# Evidence for J and H-band excess in classical T Tauri stars and the implications for disk structure and estimated ages

**Lucas A. Cieza, Jacqueline E. Kessler-Silacci, Daniel T. Jaffe, Paul M. Harvey, and Neal J. Evans II.**

University of Texas at Austin, 1 University Station, C-1400, Austin, TX 78712-0259

**Abstract:**

We argue that classical T Tauri stars (cTTs) possess significant non-photospheric excess in the J and H bands (1.25 μm and 1.66 μm respectively). We first show that normalizing the spectral energy distributions (SEDs) of cTTs to the J-band leads to a poor fit of the optical fluxes (which are systematically overestimated), while normalizing the SEDs to the $I_C$-band (0.8 μm) produces a better fit to the optical bands and in many cases reveals the presence of a considerable excess at J and H. Near-infrared spectroscopic veiling measurements from the literature support this result. We find that J and H-band excesses correlate well with the K-band (2.2 μm) excess, and that the J-K and H-K colors of the excess emission are consistent with that of a black body at the dust sublimation temperature (~ 1500-2000 K). We propose that this near-IR excess originates at a hot inner rim, analogous to those suggested to explain the "near-IR bump" in the SEDs of Herbig Ae/Be stars. To test our hypothesis, we use the model presented by Dullemond et al. (2001) to fit the photometry data between 0.5 μm and 24 μm of 10 cTTs associated with the Chamaeleon II molecular cloud. We find that simple models that include luminosities calculated from $I_C$-band magnitudes and an inner rim may account for the reported J and H-band excesses. The models that best fit the data are those where the inner radius of the disk is larger than expected for a rim in thermal equilibrium with the photospheric radiation field alone. In particular, we find that large inner rims are necessary to account for the mid infrared fluxes (3.6 – 8.0 μm) obtained by the *Spitzer* Space Telescope (*Spitzer*). The large radius could be explained if, as proposed by D'Alessio et al. (2003), the UV radiation from the accretion shock significantly affects the sizes of the inner holes in disks around cTTs. Finally, we argue that deriving the stellar luminosities of cTTs by making bolometric corrections to the J-band fluxes, which is the "standard" procedure for obtaining cTTs luminosities, systematically overestimates these luminosities. The overestimated luminosities translate into underestimated ages when the stars are placed in the H-R diagram. Thus, the results presented herein have important implications for the dissipation timescale of inner accretion disks.

**1. Introduction:**

Some of the first near-infrared observations of pre-main sequence (PMS) stars revealed ~2-5 μm fluxes well above predicted photospheric values ( e.g., Mendoza 1966 and 1968). This near-IR excess was soon recognized as evidence of heated dust in circumstellar disks, well before disks were physically resolved at millimeter wavelengths ( e.g., Kitamura et al. 1996), and later in the near-IR by interferometric observations ( e.g., Akeson et al. 2000). Over the last decades, evidence has accumulated supporting the idea that circumstellar disks are the birthplaces of planets since the disk masses, sizes, and compositions are consistent with the presumed pre-planetary solar nebula ( e.g., Hillenbrand 2003). For this reason, the study of the structure and evolution of circumstellar disks has become crucial to our understanding of the formation of planetary systems, a field that has been greatly stimulated by the newly discovered exoplanets orbiting nearby main sequence stars (e.g., Marcy & Buttler 1998).

Classical T Tauri stars (cTTs), which are low mass PMS stars still accreting circumstellar material, have large ultraviolet (UV), optical, and infrared (IR) excesses that can dominate the photospheric emission at many wavelengths ( e.g., Hartigan et al. 1991). These excesses are produced by a variety of mechanisms, all of which are associated with the presence of a disk around the young central source. The current paradigm for the structure of circumstellar disks associated with T Tauri stars ( e.g., Hartmann 1998) describes the observed SEDs in terms of the superposition of several components: the star itself, a flared disk, possibly with a hot atmosphere, and magnetospheric accretion columns through which the circumstellar material is channeled onto the surface of the star. Each component contributes a different percentage of the total flux of the system at different wavelengths, and it is usually difficult to disentangle each contribution since degeneracies arise among many of the parameters that go into modeling the SEDs (Chiang et al. 2001). The broad wavelength range of the non-photospheric emission and the frequent presence of significant circumstellar reddening in cTTs makes it difficult to find a wavelength at which to obtain photometry of the star itself from which to estimate the stellar luminosity. The method used most frequently to derive bolometric luminosity of the stellar photosphere includes applying a bolometric correction to a single-band measurement corrected for extinction ( e.g., Kenyon & Hartmann 1995 and Hartigan et al. 1994). It is usually argued that the J-band, at 1.25 μm, is the best representation of the photospheric emission. The ratio of the radiation from the photosphere to that from the hot accretion shock (UV-excess) reaches a maximum here, while the effects of extinction are less important than at shorter wavelengths and the emission from the circumstellar dust is less prominent than at longer IR wavelengths. A detailed discussion supporting this argument can be found in Kenyon & Hartmann (1990). They investigate from a theoretical point of view, the change in apparent luminosity of K1-M1 cTTs due to several effects: the occultation of the star by the disk, the accretion and reprocessing luminosity of the disk, and the radiation from the boundary layer between the disk and the stellar photosphere. They conclude that the emission from the hot boundary layer will contaminate the photospheric emission at wavelengths < 0.8 μm, while the disk emission will affect wavelengths > 2 μm; and therefore, that the I and J-band are the best representations of the true stellar fluxes. The same is true for models that replace the boundary layers with magnetospheric accretion columns ( Johns-Krull & Valenti, 2001; Calvet & Gullbring, 1998 ). Even though the presence of significant J-band excess in cTTs has been reported in the past (e.g., Folha & Emerson, 1999), the J-band is still considered to be the best representation of the photospheric emission and is commonly used, without veiling corrections, to calculate the stellar luminosity of cTTs and to derive their ages.

Here we present additional results that suggest that classical T Tauri stars (cTTs) possess significant non-photospheric excesses in the J and H bands. In section 2, we describe our SED fitting method and show that normalizing the photospheres of cTTs to the J-band leads to a poor fit of the optical fluxes (which are systematically overestimated). We show that normalizing the SEDs to the $I_C$-band [1] produces a better fit in the optical bands, $BVR_CI_C$, and in many cases reveals the presence of considerable J and H-band excesses. In section 3, we describe near-IR veiling measurements from the literature that provide independent evidence supporting our results, and in section 4 we calculate the J-K and H-K colors of the excess emission, which are consistent with black body emission at ~1500-2000 K. In section 5, we fit the photometry data between 0.4 µm and 24 µm of 10 cTTs associated with the Chamaeleon II molecular cloud and show that the reported J-band excess can be accounted for by the emission of an inner rim at the dust sublimation temperature. Then, in section 6 we investigate the effects of the J-band excess on estimating stellar ages. Finally, our conclusions are summarized in section 7.

## 2. SED fitting

### 2.1. J and H-band excesses from SED fitting:

We were motivated to investigate the possibility of significant J and H-band excesses when trying to estimate the luminosities of a sample of 15 cTTs in the Chamaeleon II molecular cloud. The sample was taken from Hughes & Hartigan (1992), and the goal was to obtain stellar ages by placing the objects in the H-R diagram, following the "standard procedure" ( e.g., Kenyon and Hartmann, 1995, hereafter, KH95). This procedure involves applying a bolometric correction, appropriate to the spectral type of the object, to a single-band measurement corrected for extinction. According to the current paradigm, luminosities obtained from the J-band and I-band should produce similar results. This is certainly the case, to within ~5 %, when the method is applied to weak-lined T Tauri stars (see sections 2.2 and 6). However, we find that when we apply this method to cTTs, the luminosities obtained from the J-band were systematically higher, by a factor of ~1.35, than those obtained from the $I_C$-band.

[1] $I_C$ denotes the I Cousins band at 0.80 µm as defined by Bessel (1979)

In order to investigate this discrepancy, we plot the entire SEDs of the stars in the Chamealeon II sample using broad band photometry, and try to separate the photospheric contribution from the rest of the SED. Table 1 lists the fluxes used to construct these SEDs. The $BVR_CI_C$ photometry and spectral types were taken from Hughes & Hartigan (1992), the JHK values come from the 2-Micron all Sky Survey (2MASS; Kleinmann, 1992), and the mid and far IR photometry was obtained as part of the *Spitzer* Legacy Project “From Molecular Cores to Planet-forming Disks (c2d)” (Evans et al. 2003). A detailed discussion of the *Spitzer* observations is presented by Porras et al. (2005) and Young et al. (2005).

As a first step in our SED fitting approach, extinction is estimated from the $R_C$-$I_C$ color excess. As discussed in section 1, at least some cTTs are known to have important non-photospheric V-band excess emission, and we argue that J and H-band excesses are also present; therefore, of all the available colors, Rc-Ic should provide the most reliable measurement of the true photospheric colors of cTTs. Following the extinction curve provided by the Asiago database of photometric systems[2] (Fiorucci & Munari 2002), we adopt $A_V = 4.76[(R_C\text{-}I_C)\text{-}(R_C\text{-}I_C)_o]$ (for $R_V = A_{V/}/E(B\text{-}V) = 3.1$) . Where $(R\text{-}I)_o$ is the expected color of a dwarf main sequence star (from KH95) with the same spectral type as the given Chamaeleon II cTTs. Then, we calculate the extinctions for all the other bands using the relations listed in Table 2, also derived using the Asiago database of photometric systems. The expected optical and near-IR fluxes are then obtained from the $I_C$ or J-band photometry corrected for extinction and the broad-band colors of main-sequence stars taken from KH95. Similarly, the predicted stellar fluxes in the *Spitzer* bands were obtained from the *Spitzer* Science Center online tool, Stellar Pet[3], which computes the mid and far-infrared fluxes using Kurucz models (Kurucz 1993) given the K magnitude and spectral type of the star. Since cTTs are known to have K-band excess, we used the predicted K-band *photospheric* fluxes calculated as described above as the input for Stellar Pet, rather than the observed K-band fluxes. Finally, all the optical and near-IR magnitudes are converted to flux densities in units of Jansky using the zero-points listed in Table 2.

[2] http://ulisse.pd.astro.it/Astro/ADPS/

The left column of Figure 1 shows that, if the SEDs are normalized to the J-band (i.e., the de-reddened J-band flux is assumed to accurately represent the photospheric flux), the BVRI fluxes are significantly overestimated. Normalizing the SED to the $I_C$-band, as shown in the right column of Figure 1 leads to a considerably better fit of the optical bands while revealing significant J and H-band excess for many of the sources. This behavior in the SEDs is not consistent with random errors and seems to be systematic. *If* our SED fitting procedure is correct, either the J-band excess is real, or the $BVR_CI_C$ fluxes are suppressed. The facts that the J and H-band excess are accompanied by excesses at longer wavelengths and that in general the observed (extinction-corrected) optical *colors* match the expected photospheric colors, suggest that the J and H-band excesses are real. We note that the accretion shock emission can easily account for the B and V-band excesses seen in some of the SEDs in Figure 1, which are in fact expected (Hartigan et al. 1991).

[3] http://ssc.spitzer.caltech.edu/tools/starpet/

**2.2. Testing the SED fitting procedure:**

In plots such as those in Figure 1, photometric uncertainties (usually around 3% in the optical and the near-IR) are small compared to other sources of error, which include errors in the spectral types, the adopted colors, and extinction corrections. To estimate the internal errors in the SED fitting approach, we applied the same procedure to a sample of 71 weak-lined T Tauri stars (wTTs) associated with the Taurus molecular cloud. Thirty nine stars of this sample were Taurus wTTs observed by the Spitzer Space Telescope (SST) as part of the Legacy project c2d (Evans et al. 2003) and are listed in Table 3. The rest of the stars in the sample were wTTs studied by Strom et al. (1989) and are listed in Table 4. It is currently believed that the main difference between cTTs and wTTs is the presence in cTTs of an inner accretion disk (Hartmann 1998) and the accompanying phenomena: strong winds and bipolar outflows, near-IR excess, UV excess, strong H$\alpha$ emission, spectral veiling, etc. All these phenomena are directly connected to the excess radiation at near-IR and shorter wavelengths; therefore, it is reasonable to assume that the fluxes of wTTs at wavelengths shorter than ~ 2 μm are a good representation of the underlying photospheres of cTTs of the same spectral type. This assumption is not valid at wavelengths longer that ~2 μm where some wTTs also possess an IR excess (Padgett et al. 2005 and Cieza et al. 2005). Following the idea that cTTs and wTTs have similar photospheres, the difference between the observed SEDs of classical and weak-lined T Tauri stars of the same spectral type can be attributed to a non-photosperic component in the cTTs fluxes for $\lambda$ < 2 μm . Tables 3 and 4 list the broad band photometry and spectral types used to fit the SEDs of our sample of wTTs. Some of the wTTs SEDs (normalized to the $I_C$-band) are shown in Figure 2 as an illustration of the good agreement between expected and extinction-corrected observed fluxes for stars of different spectral types. The solid line indicates the expected stellar photosphere (calculated as described in section 2.1. i.e, based on expected broad band colors normalized to the Ic band) and is not a fit to the extinction-corrected data points.
The excellent agreement between the expected and extinction-corrected fluxes gives us confidence in the stellar intrinsic colors and extinction corrections that we use.

The optical photometry for the wTTs listed in Table 3 comes from Cieza et al. (2005), while the spectral types for these wTTs were taken from Herbig & Bell (1988) and Wichmann et al. (2000). The optical photometry and spectral types of the wTTs in Table 4 are taken from Strom et al. (1989). For consistency, all the JHK fluxes are from 2MASS. In the case of wTTs, we find that all bands fit noticeably better than for cTTs, and normalizing the SEDs to either the J or $I_C$-band leads to essentially the same fluxes. Figure 3a shows the J-band excess for the Taurus wTTs when the photosphere is normalized to $I_C$. We define the J-band excess, $J_x$, as $J_x = \frac{J_{obs}}{J_{\exp}} - 1$, where $J_{obs}$ and $J_{exp}$ are the extinction-corrected observed fluxes and expected fluxes respectively. The mean and the median of the $J_x$ distribution for our sample of wTTs are 0.07 and 0.06 respectively, and the standard deviation is 0.14. This is consistent with wTTs having no J-band excess. Given the large number of wTTs in our sample, we believe that the 6% deviation of the median of the distribution from 0 might reflect a small, but measurable difference between the colors of T Tauri stars and those of dwarf MS stars. Such a difference in the colors is not surprising because T Tauri stars have lower photospheric gravities than dwarf MS stars of the same spectral type. We take this difference in the mean colors into account when we calculate the J-band excess of cTTs by folding in the offset of the wTTs distribution into our calculations. Thus, for each cTTs the J-band excess is calculated as: $J_x = \frac{J_{obs}}{1.06 \cdot J_{\exp}}$ - 1. The standard deviation of the distribution of Jx for wTTs is a measurement of the errors introduced by our SED fitting procedure. These errors include: errors in the spectral types, errors in the extinction correction applied, and errors introduced by the *photospheri*c variability of wTTs (the optical and near-IR photometry correspond to different epochs). We use the standard deviation of the $J_x$ distribution for wTTs as an estimate of the 1 $\sigma$ error in our procedure when we calculate the J-band excess of cTTs. However, we caution that the UV-excess produced by the accretion shock provides an important additional source of error when our procedure is used to calculate the near-IR excess of cTTs. First, the optical veiling due to the accretion shock is likely to affect the photospheric colors and the extinctions derived from the observed color excesses. We discuss this problem in Section 2.3. Second, the optical veiling introduces a much larger variability in cTTs than in wTTs. Since we use optical and near-IR data corresponding to different epochs, the variability of cTTs willl increase the uncertainty in the near-IR excesses derived for individual sources. However, in the context of our procedure, photometric variability should only introduce random errors in the determination of the near-IR excess, and it is equally likely to increase the derived near-IR excesses as it is to decrease them. Therefore, given a large enough sample of cTTs, it should be possible to establish whether or not cTTs, as a group, present significant J and H-band excesses.

To extend our sample of cTTs we include in our analysis 44 additional cTTs associated with the Taurus-Auriga molecular complex. These objects, with $BVR_CI_CJHK$ photometry and spectral types also from Strom et al. (1989) are listed in Table 5. For consistency with the Chamaeleon II cTTs, we use the JHK photometry from 2MASS.

Figure 3b shows the distribution of $J_x$ for the sample of Taurus wTTs and the combined sample of Taurus cTTs: the 15 Chamaeleon II objects from Hughes & Hartigan (1992) plus the 44 Taurus objects from Strom et al. (1989). Defining 1 $\sigma$ and $J_x$ as above, 65 % of the cTTs have J-band excesses larger than 1 $\sigma$, 48 % larger than 2 $\sigma$, and 32 % larger than 3 $\sigma$. The mean J-band excess for the sample of cTTs, $\langle J_x \rangle$, is 0.35. Figures 3c and 3d are analogous to Figure 3b, but show the excess in the H and K-bands. The statistics of the $J_x$, $H_x$, and $K_x$ distributions for our sample of wTTs and cTTs are listed in Table 6. The second, third and fourth columns show the statistics of the distributions of J, H and K-band excess for wTTs. In all cases, the distributions are consistent with wTTs having no near-IR excess. The standard deviations listed in the fourth column are used as an estimate of the $1\sigma$ errors of our procedure. These $1\sigma$ errors are used to calculate the percentage of cTTs with excess larger than $1\sigma$, $2\sigma$, and $3\sigma$ (last 3 columns). The main conclusions that can be drawn from Figures 3b-c and Table 6 are that, for cTTs, $\langle K_x \rangle > \langle H_x \rangle > \langle J_x \rangle$, and that these mean excesses are statistically significant in all cases.

Significant K-band excesses are expected for cTTs, and have traditionally been used as a diagnostic for the presence of circumstellar disks ( e.g., Strom et al. 1989 ). However, J and H-band excesses are not expected, and are difficult to explain by using current standard models of circumstellar disks around cTTs (Chiang & Goldreich 1997, 1999). It could be argued that this discrepancy between the near-IR SEDs of cTTs and wTTs is due to the fact that, in general, the SEDs of cTTs were much more strongly corrected for extinction. In that case, an anomalous extinction law could be responsible for the mismatch between the expected and observed fluxes at different wavelengths. However, we find no significant correlation between extinction and J or H-band excess, as illustrated by Figure 4. We have tested the effect of the extinction further by using a different extinction law, characterized by $R_V$=5.0 (See Table 2), to correct the stellar fluxes. Since the amplitudes of the observed J-band excesses are smaller than those of the H-band, and since the J-band is more affected by extinction, we concentrate our analysis on the result at the J-band. From the extinction relations listed in Table 2, we find that $E(J-I_C)_{Rv=3.1}$ = 0.32$A_V$ and $E(J-I_C)_{Rv=5.0}$=0.36$A_V$. This implies that in the context of our SED fitting approach, going from an extinction curve with $R_V$=3.1 to shallower extinction curve with $R_V$=5.0, would change the observed J-band fluxes by

$$\Delta J = -J_{excess} = E(J - I_C)_{R_V=3.1} - E(J - I_C)_{R_V=5.0} = -0.04 \times A_V$$

Thus, an extinction law characterized by $R_V$=5.0, could only account for the J-band excesses of the handful of objects to the left of the line $J_x/A_V$=0.04 drawn in Figure 4 (left panel), all of which have insignificant J-band excesses (< 1 $\sigma$). The same argument applies to the H-band excesses. An extinction law characterized by $R_V$=5.0 could only account for the H-band excesses for the objects also to the left of the line $H_x/A_V$=0.04 drawn in Figure 4 (right panel). Thus, we conclude that our results regarding J and H-band excesses are not significantly affected by the choice of extinction law. This very weak dependence of our results on the extinction law is due to two factors: First, since we estimate the extinction from the $R_C$-Ic color excess, the difference in extinction obtained from the two different extinction laws is less than 5%. Second, in order to estimate J-band excesses, we are effectively comparing observed extinction-corrected J-$I_C$ colors to expected J-$I_C$ colors. Since different extinction curves start to converge at these wavelengths, they predict very similar J-$I_C$ color changes for a given $A_V$.

We have also investigated the propagation of the spectral type uncertainties into the derived near-IR excesses. We find that adopting spectral types that are one sub-class later (i.e,. lower effective temperatures) than the spectral types tabulated in Tables 1 and 4 for *every* cTTs in our sample leads to an increase of ~ 0.1 and ~0.15 in the calculated *mean* J and H-band excesses respectively with respect to the excesses shown in Table 5. Similarly, adopting spectral types that are one sub-class earlier (i.e., higher effective temperature) than those shown in Tables 1 and 5, leads to a decrease of ~0.1 and ~0.15 in the mean J and H-band excesses respectively. We conclude that, unless we have *systematically* underestimated the stellar temperatures by 3 spectral type sub-classes, the J-band excess can not be attributed to uncertainties in the spectral types. To account for the H-band excesses, an even larger *systematic* error in spectral types is needed.

**2.3. Revisiting initial assumptions**

In order to estimate the J, H, and K-band excesses in section 2.2, we implicitly made two assumptions that are necessary to estimate the extinction and normalize the expected fluxes to a particular band. Namely, we assumed (1) that the observed extinction-corrected $I_C$-$R_C$ colors of cTTs correspond to photospheric colors, and (2) that the extinction-corrected Ic-band fluxes of cTTs are an accurate representation of the underlying photospheres. Then, we calculated the J, H, and K-band excesses by computing the $\frac{F_{observed}}{F_{expected}}$ flux ratios, where $F$ stands for the J, H or K- band fluxes. In the context of our procedure, this is equivalent to calculating $I_C$-J, $I_C$-H, and $I_C$-K color excesses according to $m_{color-excess} = (I_{Cobs} - m_{obs}) - (I_{C\exp} - m_{\exp})$, where *m* stands for J,H,K magnitudes. If both assumptions (1) and (2) are correct, then the color excess accurately measures the true non photospheric excess, $m_{excess}$. However, since the emission from the accretion shock and the inner disk will also contribute to the $I_C$ and R-band total fluxes, these two assumptions are only approximations. In order to test their validity, we take the case of a M0 cTTs, the most common type of star in our sample, with a J-band excess equal to the mean J-band excess reported in section 2.2 ($r_J = 0.35$), and a V-band excess equal to the mean V-band excess ($r_V$=0.60) reported by Gullbring et al. (1998) and Hartmann & Kenyon (1990) for a sub-sample of the objects in Table 5. Using the mean colors of main-sequence stars from KH95 and assuming that the emission from the accretion shock and the inner rim can be characterized as black body emission at 10,000 K and 1,700 K, respectively, we derived the expected veiling at the $I_C$ and $R_C$-bands shown in Table 7. The last column shows the total change in apparent magnitude due to the veiling produced by the accretion shock (second column) and the rim (third column) emission. The values are for a M0 star with typical J and V-band veiling of 0.60 and 0.35 respectively. We find that the $R_C$ and $I_C$-bands contain a non-photospheric contribution of 28% and 19% respectively [4]. Since we normalized the photosphere to the $I_C$-band, a zero color excess, $m_{color-excess} = 0$ , for a given band, would actually imply: $m_{excess} = r_{Ic} = 0.19$ ( i.e., it seems that we *underestimate* the J, H and K-band excesses by 0.19). However, there is another effect that compensates for the fact that we ignore the veiling at $I_C$. Since $r_{Ic} = 0.19$ and $r_{Rc} = 0.28$, the $I_C$-$R_C$ colors of the stellar photosphere appear *bluer* by 0.09 magnitudes, and we *underestimate* the extinction, $A_V$, by 0.56 magnitudes. If the extinction is underestimated, the shortest wavelengths of the SED are under-compensated with respect to the longer wavelengths, and an artificial color excess is produced. Using the extinction relations from Table 2, we convert the underestimated extinctions into the apparent color excesses shown in Table 8. The second column lists the amount by which the extinction is *underestimated* due to the change in $R_C$-$I_C$ colors produced by the veiling listed in Table 7. The third column shows the amount by which the color excess is *overestimated* due to the underestimated extinctions, $\Delta(I_C\text{-}m)_{Av}$. The fourth column shows the net effect of ignoring both $r_{Ic}$ and $r_{Rc}$ on the apparent excesses at the $BVR_CI_CJHK$ bands (for a M0 star with $r_V$=0.60 and $r_J$=0.35). For the J, H and K-bands, the end result is that $m_{excess} \approx m_{color-excess}$ to within 5%, which was the original assumption.

[4] The superposition of two different sources of excess emission, one hotter and one cooler than the stellar photosphere, provide a simple explanation for the observations showing that the veiling in cTTs "flattens out" in the red part of their optical spectra (e.g., Basri & Batalha, 1990 and White & Hillenbrand, 2004).

Also, we find that the change in the apparent $I_C$ magnitude due to the veiling, $\Delta m_{Ic}$, is well compensated by the underestimation in extinction in that band, $\Delta A_{Ic}$. In fact, $\Delta A_{Ic} - \Delta m_{Ic}$ ~ 0.05 mag, which implies that assuming no $I_C$ and $R_C$ excess only affects the apparent luminosity by ~5 %.

The last column in Table 8 also shows that, in this example, we underestimate the $R_C$ and $I_C$-band excesses by exactly the same amount as the assumed veiling (0.28 and 0.19 magnitudes, respectively). Similarly we underestimated the V-excess by 0.37 magnitudes which is equivalent to underestimating the veiling by 0.4. Since the assumed V-band veiling was 0.6, this means that the SED fitting approach will typically reveal only ~30% of the V-band excess due to the accretion shock. This compensating effect of the underestimated extinction on the optical excess explains why the optical SEDs shown in the left panel of Figure 1 match the expected photospheres so well, even though excess emission is likely to be present at all wavelengths.

This interplay between the UV-excess and the apparent extinction prevents us from obtaining the V-band or B-band excess from the SED and improving the fit recursively by taking into account the effect of the veiling on the apparent colors.

## 2.4. Comparison with previous works

We arrive at the conclusion that cTTs possess significant J and H-band excesses by analyzing photometric data that are available in the literature. Thus, we were motivated to compare our procedure and assumptions against those found in the original papers from which most of the data were taken (i.e., Strom et al. 1989 and Hughes & Hartigan 1992). We also compare our procedure with that followed by Meyer et al. (1997), who present a detailed analysis of the near-IR colors of cTTs.

Strom et al. (1989, S89 hereafter), present SEDs for 16 of the T Tauri stars in our Table 4 and Table 5. Their SEDs are normalized to the $R_C$-band, and as a photospheric model, they use SEDs of dwarf stars of a spectral type corresponding to that of the T Tauri stars. Even though the presence of significant J and H-band excesses is not heavily emphasized by S89, these excesses are clearly seen in most of their SEDs. In fact, S89 mentions that in some cases the spectral energy distribution of the *excess emission* can be characterized as black body emission at a temperature of T ~ 2000-2500 K and suggest that the most likely origin of this emission is the inner edge of the disk at the dust sublimation temperature. This conclusion is one of the main results presented herein (see section 4.2), but it has been for the most part neglected by subsequent literature. Possibly, the large uncertainties in their procedure and the high temperatures derived from the excess emission prevented S89 from making a stronger case for the presence of significant J and H-band excesses. Several factors may have contributed to a larger uncertainty in the SED fitting procedure used by S89 when compared to our procedure. First, S89 calculate the extinction from the (V-$R_C$) color excess which is more sensitive to the veiling produced by the accretion shock luminosity and to the extinction law than is the ($R_C$-$I_C$) color excess we use. Second, they use intrinsic colors from Johnson (1964) which are on the Johnson system, not on the Cousins system as the observations they report. The transformation between photometric systems introduces an additional source of error. Finally, S89 use J, H and K-band photometry compiled from the literature, while we use the 2MASS catalog which provides a more uniform data set.

Hughes & Hartigan (1992, HH92 hereafter) present SEDs for all the objects shown in Figure 1. They normalize the SEDs to the J-band (i.e., they assume zero J-band excess as we do for the SEDs shown in the right column of Figure 1.). However, their SEDs do not show the clear systematic underestimation of the optical fluxes seen in our SEDs when they are normalized to the J-band. It is likely that the systematic underestimation of the optical fluxes is masked by the large uncertainties in their procedure. HH92 calculate the extinction, as we do, from the ($R_C$-$I_C$) color excess of the objects, but do not specify the extinction law used. They adopt intrinsic colors taken from Bessel (1979) who only reports intrinsic colors for a very limited set of spectral types (i.e., F5, G0, G6, K2, K4, K7, M2); therefore, they probably had to interpolate in order to obtain intrinsic colors for stars of intermediate spectral types. Also, and more importantly, they use a black body curve as the stellar model, which provides only a very rough approximation of the photospheric fluxes.

Meyer et al. (1987, M97 hereafter), follow a procedure very similar to ours in order to calculate the near-IR excess of cTTs. However, they made the crucial assumption that the non-photospheric contribution to the J-band flux comes exclusively from tail of the UV-excess produced by the accretion shock (i.e., there is no contribution from the disk). With this assumption, they estimate that the J-band veiling is 10% of the V-band veiling and calculate the J-band excess from the V-band veiling values provided by Hartigan et al. (1995). They find that the mean of the J-band veiling calculated in this way is $<J_x>$ ~ 0.0. M97 analyze the same sample of cTTs reported by S89 (Table 4 in this paper), but they use the original near-IR fluxes provided by S89 rather than the 2MASS fluxes used by us. They calculate the extinction from the ($R_C$-$I_C$) color excess and use extinction corrections *identical* to ours (i.e., E($R_C$-$I_C$)=0.21 $A_V$, E(J-H) = 0.11$A_V$, and E(H-K)=0.06$A_V$). However, M97 adopt intrinsic colors from Bessel (1979), which has the limitations mentioned above. With this assumption that $<J_x>$ ~ 0.0, they estimate the H and K-band excess from the J-H and J-K color excesses. M97 find a *median* H and K-band excess of 0.2 and 0.6 respectively, but caution that the reported values are only lower limits because of the assumption of zero J-band excess. In fact, they state that if they normalize the photospheres to the $I_C$-band, the calculated mean J-band excess becomes 0.23. In section 2.2 we found *median* J-band , H-band, and K-band excesses of 0.28, 0.54, 1.1, respectively, for our combined sample of Chameleon and Taurus cTTs. We conclude that, once the M97 excesses are corrected for the assumption of zero J-band excess (by adding $<J_x>$ ~ 0.25 to the $<H_x>$ and $<K_x>$ excesses ), their values agree well with our calculated J,H, and K-band excesses. K-band veiling measurements from the literature support larger K-band excess values than the 0.6 reported by M97 (closer to our 1.45 calculated *mean* value). Folha et al. (1999) obtain a mean K-band veiling, $<r_K>$ ~ 1.3 for a sample of 30 Taurus cTTs and Doppmann et al. (2003) calculates $<r_K>$ ~ 2.0 for a sample of 10 Ophiuchus cTTs, while Muzerolle et al. (2003) finds $<r_K>$ ~1.2 for a sample of 9 Taurus cTTs. In the next section we discuss more spectroscopic veiling measurements that support our conclusion that classical T Tauri stars present significant J and H-band excesses.

## 3. Spectroscopic evidence for J-band excess

In order to test our results from the previous section indicating the presence of significant J and H-band excesses, we analyze a sub-sample of the cTTs in the Taurus-Auriga complex with spectroscopic J-band veiling measurements available in the literature. These measurements provide a test that is independent of any assumptions regarding reddening, extinction or broad-band colors. Spectral veiling, $r_\lambda$, is defined as the ratio of any non-photospheric flux to the photospheric flux at a given wavelength, $\lambda$. This excess flux is usually estimated by comparing the equivalent widths of the lines of the program objects to those of unveiled stars used as templates, or to synthetic models.

Perhaps because J and H-band excesses are not expected, we find no H-band veiling measurements of cTTs in the literature, and only a few works reporting J-band measurements. However, Folha & Emerson (1999), FE99 hereafter, report J-band veiling measurements for 45 cTTs, 33 of which have $BVR_CI_CJHK$ photometry from Strom et al. (1989). This data set provides a sample to test directly our results from previous sections. The FE99 veiling measurements, listed in Table 5 as "$r_J$ Spectra", were obtained from high resolution spectra (R ~20,500) around the Pa $\beta$ line (1.28215 µm) using Main Sequence dwarfs of similar spectral types as templates. Also listed in Table 5 are the J-band excesses calculated using our SED fitting approach ($J_x$ SED). The tabulated $J_x$ SED values, defined as in section 2.2, can be directly compared with the veiling values obtained by FE99. Figure 5 shows the J-band excesses obtained by the two different methods. Since the data points cluster on the upper-right quadrant of the figure, both methods show clear evidence of J-band excess for cTTs as a group. We note that the average and range of the J-band excesses measured by these two different methods are in good agreement, even though the agreement for individual objects is relatively poor. The spectroscopic and photometric data correspond to different epochs, however, and variability might be responsible for some of the scatter. Comparison of the J-band magnitudes reported by Strom et al. (1989) and those from 2MASS show an average difference of ~ 0.2 magnitudes and a maximum deviation of up to a factor of 2 in flux, but no systematic variation. Also, The 1 $\sigma$ error bars shown for $J_x$ (SED) correspond to the standard deviation of the $J_x$ in wTTs listed in Table 5, and do not include the errors introduced by the interplay between the UV excess and the apparent extinction discussed in section 2.3. These errors are difficult to quantify, but are likely to weaken the expected correlation between $J_x$ (SED) and $J_x$ (Spectroscopy).

Other somewhat less direct, but still compelling, evidence for J-band excess is presented by Doppmann et al. (2003). They obtained K-band veiling measurements of 10 cTTs associated with the Rho Ophiuchus dark cloud from high resolution spectra (R = 50,000) centered around 2.207 μm. In this case, the veiling is obtained using spectral synthesis models as templates. They compare stellar luminosities from dereddened K-band magnitudes corrected for veiling against luminosities derived from dereddened J-band magnitudes assuming zero J-band veiling. They find that the J-band luminosities are systematically higher by a factor ~ 2. This implies an average J-band veiling of ~1, which is higher than the average J-band veiling of ~ 0.6 found by FE99 for the Taurus cTTs, and the average J-band excesses of ~ 0.4 from the SED fitting obtained in this work for the cTTs in Taurus and Chamaeleon II. However, the cTTs in the Doppmann Rho Oph sample were selected based on their large K-band luminosities. Since K-band excess usually dominates the photosphere (they found $\langle K_x \rangle = 2.0$ ), the sample is probably biased toward large K-band and J-band excesses.

SED fitting and spectral veiling measurements independently provide compelling, but not conclusive, evidence for the existence of J-band excesses in cTTs. The combination of these two independent lines of evidence, however, provides a very strong case for the presence of significant non-photospheric J and H-band excesses in cTTs. The existence of a J-band excess has important implications for the study of the structure and evolution of cTTs disks and should be investigated further.

## 4. The physical origin of the near–IR excess

### 4.1. J and H-band excess vs. K-band and V-band excesses.

In order to explore the nature of the J and H-band excesses, we investigate their correlation with the two known sources of non-photospheric radiation: the accretion shock and the disk emission. If the J and H-band excesses are related to the accretion shock, they should correlate with optical veiling, $r_V$, as measured by spectral veiling (i.e, $r_J \sim 0.1\ r_V$ for late K and early M stars). Figure 6 shows our J-band excess measurements versus the $r_V$ from Gullbring et al. (1998). We do not find any strong correlation with this small data set, but clearly, $r_J >> 0.1\ r_V$, instead of $r_J \sim 0.1\ r_V$, as would be expected if both originated directly at the accretion shock (Hartigan et al. 1995). In addition, the emission from the accretion shock should be negligible at the H-band, but we find that $\langle Hexc \rangle > \langle Jexc \rangle$. Thus, we discard this explanation.

If the J and H-band excesses come from the circumstellar disk itself, one might expect them to correlate with the excess at longer wavelengths. Figure 7 shows our calculated K-band excess vs. J and H-band excesses (left and right panel respectively) for both the Chamaeleon II and Taurus cTTs from Tables 1 and 5. The Spearman's ranks of these correlations are 0.65 and 0.92 with probabilities of being drawn from a random distribution of $1.5 \times 10^{-8}$ and $6.3 \times 10^{-26}$ respectively. These are robust correlations, and they strongly suggest that the J, H and K-band excesses have a common source.

**4.2. The color temperature of the near-IR excess.**

If J, H and K-band excesses have a common source, and this source is optically thick, then its characteristic temperature can be estimated from the J-K and H-K colors of the excesses, or equivalently the ratio of the J to K and H to K excess fluxes. Our SED fitting approach allows a straightforward calculation of the J-K and H-K colors of the excess. Following the discussion in section 2.2, we obtain: $J_{EXC} \approx J_{obs} - 1.06 \cdot J_{\exp}$ , $H_{EXC} \approx H_{obs} - 1.03 \cdot H_{\exp}$ , and $K_{EXC} \approx K_{obs} - 1.06 \cdot K_{\exp}$ . Where, $J_{EXC}$, $H_{EXC}$ and $K_{EXC}$, are the absolute J,H, and K-band excesses *fluxes* in Jy, as opposed to the dimensionless excess $J_x$, $H_{exc,}$ and $K_{exc}$ discussed so far. Figure 8 shows $J_{EXC}$ vs. $K_{EXC}$ and $H_{EXC}$ vs. $K_{EXC}$ in units of flux of the expected stellar photospheres at 2.2 μm. The *flux* ratios $\frac{J_{EXC}}{K_{EXC}}$ and $\frac{H_{EXC}}{K_{EXC}}$ shown in Figure 8 are both consistent with black body emission at a relatively narrow range of temperatures, T ~ $1750 \pm 250$ K. The right panel of Figure 8 reveals a tighter correlation than the left panel. This is expected however, because the percentage error in $J_{EXC}$ is about twice the percentage error in $H_{EXC}$. The 1 $\sigma$ error bars shown in the top-left of both panels correspond to the standard deviation of wTTs listed in Table 6. For the reasons discussed in section 2.3, the actual error bars are probably larger, suggesting that uncertainties in our procedure are responsible for a significant fraction of the scatter in the observed excesses.

Depending on the density and composition, the sublimation temperature of dust grains is also ~ 1500-2000 K ( e.g., Pollack et al 1994). Thus, we argue that the near-infrared excess of T Tauri stars is produced at the inner edge of the disk whose temperature is set by the dust sublimation temperature.

**4.3. The color Temperature of the IRAC excesses**.

Following the procedure outlined in section 4.2., we obtain the color temperature of the mid-IR excess of the cTTS from the Chamaeleon II sample (Table 1) by computing the ratios of the flux excesses at the IRAC[5] -1 (3.6 µm) and IRAC-3 bands (5.8 µm) (Porras et al. 2005). At these wavelengths, the excess emission largely dominates over the photospheric emission, and the uncertainties in the expected fluxes are likely to dominate the errors in deriving color temperatures. Figure 9 shows that the color temperatures of the IRAC excess are T ~ 1400 $\pm$ 200 K). This temperature is similar to the black body temperature derived by Muzerolle et al. (2003) (T ~ 1400 K) in which they used high resolution spectroscopy of three spectral regions between 2.1 and 4.8 µm to probe the shape of the excess emission of 9 cTTs. These temperatures are significantly lower than those obtained from the near-IR colors of the excess (T ~1750 $\pm$ 250 K). We discuss a possible explanation for this difference in the following section.

[5]The Infra-Red Array Camera on the *Spitzer* Space Telescope

### 4.4. The inner disks of PMS stars

Herbig Ae/Be stars (Herbig 1960) are pre-main sequence intermediate mass (mass $\geq$ 2 $M_{\odot}$) stars analogous to cTTs (mass < 2 $M_{\odot}$). The mid- and far-IR regions of the SEDs of Herbig Ae/Be stars are well fitted by standard models of passive flared disks ( e.g., CG97 & CG99); however, these models fail to explain the near-IR excess, known as the "near-IR bump," observed in most Herbig Ae/Be stars. According to these simple models, the disk flares outward at large radii due to the vertical hydrostatic equilibrium, but is physically thin near the star. Thus, the grazing angle of the incident radiation is small at small radii, and the inner disk is heated very inefficiently and extends to a few stellar radii before reaching the dust sublimation temperature. When the vertical structure of the inner disk is taken into account (Natta et al. 2001 and Dullemond et al. 2001), the very inner edge of the disk, which becomes an inner rim, is irradiated normal to the surface and heated more efficiently. Thus, the dust sublimation radius moves farther away from the star. Farther away from the star, gravity becomes weaker, and the disk's scale height increases. The effect is that the surface area of the region emitting at the dust sublimation temperature becomes much larger than that predicted by simple standard disk models. With this modification, the radiation from the inner rim can account for the observed near-IR bump. Naturally, we investigate the possibility of an analogous inner rim in cTTs to explain the observed 2MASS and IRAC excesses. The idea that cTTs might present inner rims analogous to those of Herbig Be/Ae stars has already been proposed by Muzerolle et al. (2003) based on the black body shape of their 2.1-4.8 μm excess emission and by Allen et al. (2004) based on the IRAC colors of cTTs in young clusters.

For simplicity, current circumestellar disks models usually adopt a single dust sublimation temperature. Natta el al. (2001) assume a dust sublimation temperature of 1700 K, while Dullemond et al. (2001) adopt a temperature of 1500 K to fit their models. However, recent detailed models of the shape of the inner rim (Isella & Natta 2005) show that, when the dependence of the dust sublimation temperature on gas density is taken into account, the inner rim becomes rounded, and its surface has a vertical temperature gradient which is several hundreds of K wide. Such an inner rim would present a hotter color temperature at shorter wavelengths, and a cooler color temperature at longer wavelengths, and could help to explain the discrepancy between the color temperatures we derived from the 2MASS and IRAC observations (section 4.2. and 4.3., respectively).

## 5. Disk modeling and implication of the near-IR excess for disk structure

In order to quantify the contribution from the inner rim to the total flux at different wavelengths, we model the SED of 10 cTTs from the Chamaeleon II sample (Table 1) using the disk model presented by Dullemond et al. (2001). The model is based on the flared disk model of Chiang & Goldreich (1997 and 1999) and includes a disk with three distinct components: a cool disk interior, a warm surface layer, and a hot inner rim located at the dust sublimation radius. The main parameters of the models are listed in Table 9. For all models, we assume a single dust sublimation temperature of 1400 K, corresponding to the typical color temperature of the IRAC excess found in section 4.3. To estimate the stellar effective temperatures ($T_{eff}$), we adopt the spectral type-$T_{eff}$ relations from KH95. The stellar luminosities are obtained from the extinction-corrected $I_C$-band (0.8 µm) magnitudes and the bolometric corrections, appropriate for the spectral type, from Hartigan et al. (1994). Following Hughes & Hartigan (1992), we adopted a distance of 200 pc for all the objects. Finally, the stellar masses are estimated using the evolutionary tracks presented by Siess et al. (2000). For our objects, we find that Siess et al. models yield masses that are intermediate between those derived from the models by D'Antona et al. (1998) and those obtained from the models by Baraffe et al. (1998). The photospheric luminosity of the star, the stellar mass, and the dust sublimation temperature determine the radius and scale height of the rim. This predicted inner rim is labeled "Inner rim A" on the disk models in Figure 10. We find that the models for most of the stars systematically underestimate the near and mid-IR excesses. However, a good fit can be obtained simply by scaling the contribution from the predicted inner rim by a factor, $\Omega$, that ranges from ~1 to ~7. This "scaled-up" rim is labeled as "Inner rim B" in Figure 10. We interpret this result as an indication that the *area* of the inner rim is larger, by a factor of $\Omega$, than predicted by the models. i.e. $\Omega = \frac{A_{rimB}}{A_{rimA}}$, where $A_{rim}$ is the area of the rim. Since $A_{rim} \propto R_{rim} \cdot H_{rm}$, and according to our adopted model , $R_{rim} \propto H_{rim}^{2/3}$ , where $R_{rim}$ and $H_{rim}$ are the radius and the scale height of the rim, then, $A_{rim} \propto R_{rim}^{5/2}$. Thus, the radius of the inner rim B can be calculated as $R_{rimB} = R_{rimA} \cdot \Omega^{2/5}$

The masses of the disk models shown in Figure 10 were adjusted to try to match the observed 24 µm fluxes. The adopted disk masses range from $5\text{x}10^{-2}$ to $5\text{x}10^{-4}$ solar masses. In all cases, the out disk radius was set to 400 AU, and the disk's surface density, $\Sigma$, is given by $\Sigma(R)(\text{g/cm}^{-2}) = 2 \times 10^3 (R/\text{AU})^{-2}$. But, since our simple approach of scaling the inner rim does not take into account the effects of the modifications in the inner disk on the disk structure at larger radii, we do not try to constrain the physical parameters of the outer disks. However, we keep the outer disk models in the SEDs shown in Figure 10 *only* to show that the 2MASS and IRAC fluxes are completely dominated by the emission from the inner rim with very minor contributions from the rest of the disk

The fact that the energy eradiated (i.e., the area under the curve in Figure 10) by rim B is larger than that eradiated by rim A suggests that the inner rim is powered by more than the stellar photosphere. We argue that the most likely "source of missing energy" is the UV emission from the accretion shock produced as material from the star is channeled onto the stellar surface. The accretion shock emission has already been recognized by D'Alessio et al. (2003) as an important heating source of the inner disks of cTTs. Unfortunately, as discussed in section 2.3, it is very difficult to estimate the UV-excess from the SED alone, and it needs to be obtained independently , e.g., from UV spectroscopy. However, most accretion luminosity estimates based on optical spectroscopy involve an extinction correction. Gullbring et al. (1998) estimate accretion luminosities for a sample of cTTs from UV spectroscopy and compare their results with those presented by Hartigan et al. (1995), for the same sample of stars, following a similar method. The accretion luminosities derived by these two groups systematically differ by up to an order of magnitude. According to Gullbring et al. (1998), most of the discrepancy can be traced back to a large systematic difference in the extinction corrections. The large variability typical of the UV excess makes it even harder to obtain an accurate estimate of the accretion luminosity unless the observations involved in the analysis are made simultaneously. An estimate of the UV-excess is necessary to test whether the energy from the accretion shock luminosity is enough to account for the observed mid-IR excesses seen in the Chamaeleon II objects; however, for the reasons mentioned above, we leave such a test for future work.

We note that the degeneracy between the UV-excess and the extinction can eventually be disentangled by measuring the veiling at the wavelengths corresponding to the $BVR_CI_C$ band passes using high resolution spectroscopy from 0.4 to 0.9 μm and obtaining *simultaneous* optical photometry. With that information, the $R_C$-$I_C$ colors can be corrected for veiling in order to estimate the extinction more accurately, and the UV excess can be estimated directly from the B-band veiling or the U photometry corrected for extinction. We plan to follow that procedure in a follow-up paper in order to study self-consistently the effect of the UV-excess on the SEDs of cTTs at near-IR and Spitzer wavelengths. However, even without the veiling information from spectroscopy, we do find indirect evidence that supports the idea that the UV excess significantly affects the sizes of the inner holes in disks around cTTs. First, if the inner rim is larger than expected because it is significantly powered by accretion shock luminosity, a correlation between the K-band excess and the accretion luminosity is expected. For a sub-sample of the Taurus cTTs, we use the accretion luminosities, derived from UV photometry and spectroscopy, from Muzerolle et al. (1998) to investigate the correlation between K-band excesses and accretion luminosity. This correlation is evident in Figure 11, which also shows that, for some cTTs, accretion shock luminosity can dominate the stellar luminosity. The Spearman's rank of the correlation between K-band excesses and accretion luminosity is 0.81 with a probability of being drawn from a random distribution of $1.01 \times 10^{-6}$. A similar correlation between accretion luminosity and near-IR excess has been reported by Muzerolle et al. (2003) for a smaller sample or cTTs. Also, D'Alessio et al. (2003) demonstrate that including the UV radiation in the circumstellar disk models can significantly increase the size of the inner hole. In particular, they find that, when the UV excess is included, the dust sublimation radius of the "continuum star" DG Tau (0.2 AU) is ~3 times larger than the radius inferred when neglecting the UV excess emission (0.07 AU), and is in good agreement with the inner radius derived from K-band interferometric observation of DG Tau (Colavita et al. 2003). For our objects, $\frac{R_{rimB}}{R_{rimA}} \leq 2$; thus we conclude that the UV-excess from the accretion shock could in principle account for the sizes of all the inner rims reported herein.

We were motivated to investigate the possibility of large inner rims in cTTs while trying to find an explanation for the J and H-band excesses calculated in section 2. However, we emphasize that our results from the IRAC bands, which suggest the presence of large inner rims, are independent of any assumptions made about the presence of J or H-band excesses. In section 2.2, we found that the J-band excess is at the ~35 % level. Using the J-band to obtain the photospheric luminosity, rather than the $I_C$-band, increases the expected IRAC fluxes only by ~35%. But in some cases, at IRAC wavelengths, the flux discrepancy between disk models with small inner rims heated only by the stellar photosphere and the observations is an order of magnitude larger than the J-band excess. This discrepancy between the models and the observed IRAC fluxes is well beyond any observational errors and uncertainties in the expected photospheric fluxes. We have followed the same procedure described in section 2.1 to calculate the IRAC excesses of a large sample of wTTs (Cieza et al. 2005). For wTTs showing no IR-excess, the expected photospheric fluxes agree with the observed fluxes to within ~5%. Since the existence of large inner rims in cTTs is also supported by interferometric observations (Colavita et al. 2003)., and its presence could account for both the IRAC and 2MASS excesses, it is tempting to conclude that the J and H-band excesses calculated in section 2.2 are *mainly*, even if not exclusively, produced by the tail of the inner disk emission.

## 6. Implications of the J-band excess for stellar ages and disk evolution

The presence of significant J-band and H-band excess has important implications not only for the structure of circumstellar disks, but also for estimations of stellar ages. Since cTTs are usually placed in the H-R diagram using luminosities derived from the J-band (e.g., KH95 and Hartigan et al. 1994), a systematic error in the J-band luminosities translates into a systematic error in the derived ages. In order to investigate the effect of the J-band excess on the derived luminosities and ages, we calculate the luminosities of our entire sample of cTTs and wTTs from the extinction-corrected $I_C$ and J-band magnitudes and bolometric corrections appropriate for the spectral types from Hartigan et al. (1994), and then compare the results. We adopted distances of 140 pc and 200 pc for objects in Taurus and Chameleon II, respectively (Kenyon et al. 1994, and Hughes and Hartigan, 1992). For our sample of 59 cTTs (Tables 1 and 5), we find that luminosities derived from the J-band are systematically higher by a factor of ~ 1.35 on average with respect to luminosities obtained from the $I_C$-band. However, for wTTs, we find no systematic difference between the two methods. This systematic difference in the luminosities obtained for cTTs is a direct consequence of the J-$I_C$ color excesses reported in section 2.2; therefore, the uncertainties in the J-band excess determination propagate directly into the uncertainties in the luminosity *difference* between luminosities derived from the I-band and those derived from the J-band. In section 2.3, we conclude that these color excesses are a good measurement of the non-photospheric J-band contributions (i.e., $J_{excess} \approx J_{color-excess}$). Thus, we believe that the photospheric luminosities obtained from the $I_C$-band are more accurate than those obtained from the J-band. As discussed in Section 5, if optical spectroscopic veiling measurements were available, this conclusion could be tested by combining photometry and spectroscopic veiling measurements at the $R_C$ and $I_C$ bands. The extinction can then be obtained from the veiling corrected $R_C$-$I_C$ colors and the $I_C$ fluxes can be corrected for extinction and veiling independently rather than assuming that the effects cancel each other.

If the luminosities obtained from the $I_C$ band are in fact more accurate that those obtained from the J-band as a general rule, the luminosities of cTTs have been systematically overestimated by most studies. Since low-mass PMS stars $(mass < 1\ M_{\odot})$ contract roughly at constant temperature, overestimated luminosities translate into underestimated ages when the stars are placed in the H-R diagram. This effect is shown in Figure 12, which plots the ages of cTTs and wTTs obtained from $I_C$-band luminosities vs. those obtained from J-band luminosities for 3 different sets of evolutionary tracks. We find that, in general, models by D'Antona et al (1998) (a) yield younger ages, models by Baraffe et al. (1998) (c) yield older ages, while models by Siess et al. (2000) (b) yield intermediate ages. In all cases, cTTs appear systematically younger when the ages are derived from the J-band luminosities instead of the $I_C$-band luminosities. Since wTTs have no J-band excess, no systematic effect is seen for their ages, and using J-band or $I_C$ luminosities yields essentially the same age. In Figure 13, we plot the age distribution of cTTs and wTTs when the stellar luminosities are estimated from the J-band (left panel) and from the $I_C$-band (right panel) using the models from Siess et al. (2000). The mean, median and standard deviation of the logarithmic age distribution (in million of years) are 0.32, 0.27, and 0.35 respectively when the ages are derives from J-band luminosities and 0.50, 0.47, and 0.37 respectively when the ages are derived from Ic-band luminosities.

The right panel of Figure 13 shows that, when the ages are derived from the $I_C$-band, the overlap of the age distribution of cTTs and wTTs increases significantly with respect to the age distributions obtained from the J-band luminosities. Most wTTs are likely to be evolutionary descendants of cTTs, since all low-mass PMS are likely to go through a cTTs phase, even if this phase is short. Strong winds and star-disk interactions are the main mechanisms through which PMS stars are believed to dissipate angular momentum; therefore, without a T Tauri phase it becomes *very* difficult to explain the angular momentum evolution of young stellar objects (Rebull et al. 2004 ). If wTTs are in fact evolutionary descendents of cTTs, a large overlap in their age distributions implies a wide distribution in the duration of the cTTs stage. In this context, the right panel of Figure 13 suggests that the inner accretion disk, the presence of which defines the cTTs phase, dissipates on a time scale that ranges from 1 to 10 Myr.

The diversity in the dissipation time-scale of the inner accretion disks might be related to the presence of sub-stellar companions or to the formation of giant planets within the disks. The presence of planets is usually invoked to account for the large inner holes (~1-10 AU wide) inferred from the SEDs of several wTTs and cTTs ( e.g., Calvet et al. 2002 and D'Alessio 2004 ). Mid and far-IR properties of a statistically significant sample of young wTTs (i.e., coeval with cTTs) are needed to test this idea. *Spitzer* observations will soon reveal the fraction of wTTs with (non-accreting) circumstellar disks as a function of age, which will help to constrain the dissipation timescale of the planet-forming region of the disk.

## 7. Summary and conclusions

1) In section 2, we showed that cTTs present significant J and H-band color excesses in addition to the well studied K-band excess. We interpreted these color excesses as evidence for non-photospheric emission.

2) In sections 4.2 and 4.3, we estimated the color temperature of the excess emission at 2MASS and IRAC wavelengths, respectively. We found that the color temperature of the excess emission is T ~ 1750 $\pm 250$ K at 2MASS wavelengths and T ~ 1400 $\pm 200$ at IRAC wavelengths. We suggested that this emission originates at an inner rim which is physically narrow but has a gradient of temperatures several hundreds of degrees wide.

3) In section 5, we modeled the SED of 10 cTTs from 0.4 to 24 μm and found that the 2MASS and IRAC fluxes are dominated by the emission from the inner rim. The models that best fit the data are those where the inner radius of the disk is larger than expected for a rim in thermal equilibrium with the stellar radiation field alone. We found that the K-band excess correlates with accretion luminosity. As proposed by D'Alessio et al. (2003), the UV radiation from the accretion shock could explain the larger than expected inner holes.

4) Finally, in section 6, we calculated stellar luminosities from the $I_C$ and J-band, and used these luminosities to estimate stellar ages from 3 different sets of evolutionary tracks. We argued that normalizing the luminosity of cTTs to the J-band systematically overestimates their luminosities. These overestimated luminosities translate into underestimated ages when the stars are placed in the H-R diagram. When the ages are derived from $I_C$-band luminosities, cTTs and wTTs show a larger age overlap with respect to ages derived from the J-band. If wTTs are descendants of cTTs, this large overlap implies a wide diversity in the duration of the cTTs phase.

Support for this work, part of the *Spitzer* Legacy Science Program, was provided by NASA through contracts 1224608 and 1256316 issued by the Jet Propulsion Laboratory, California Institute of Technology, under contract 1407. This publication makes use of data products from the Two Micron All Sky Survey, which is a joint project of the University of Massachusetts and the Infrared Processing and Analysis Center/California Institute of Technology, funded by the NASA and the NSF.

**Table 1. Chamaeleon II cTTs from Hughes & Hartigan (1992)**

| Star ID | SpT | B Mag | V mag | $R_C$ mag | $I_C$ mag | J mag | H Mag | K mag | IRAC1 mJy | IRAC2 mJy | IRAC3 mJy | IRAC4 mJy | MIPS 1 mJy |
|---|---|---|---|---|---|---|---|---|---|---|---|---|---|
| Sz 46 | M3 | 17.66 | 16.19 | 14.74 | 13.18 | 11.25 | 10.26 | 9.75 | 8.14E+01 | 7.33E+01 | 6.37E+01 | 5.55E+01 | 5.16E+01 |
| Sz 48 | M1 | 19.17 | 18.05 | 16.17 | 14.36 | 11.44 | 10.10 | 9.45 | 1.33e+02 | - | 8.97e+01 | - | 8.40e+01 |
| Sz 50 | M3 | 17.64 | 16.01 | 14.30 | 12.50 | 10.31 | 9.32 | 8.85 | 1.38e+02 | 1.37e+02 | 1.31e+02 | 1.71e+02 | 3.56e+02 |
| Sz 51 | M0 | 15.38 | 14.50 | 13.47 | 12.38 | 10.61 | 9.85 | 9.35 | 1.91e+02 | 1.64e+02 | 1.30e+02 | 1.18e+02 | 1.08e+02 |
| Sz 53 | M1 | 17.85 | 16.59 | 15.20 | 13.66 | 11.73 | 10.58 | 9.92 | 9.39e+01 | 8.72e+01 | 7.32e+01 | 7.62e+01 | 8.47e+01 |
| Sz 54 | K7 | 13.88 | 12.53 | 11.58 | 10.61 | 9.05 | 8.15 | 7.59 | 3.87e+02 | 4.05e+02 | 3.08e+02 | 2.71E+02 | 2.61e+02 |
| Sz 55 | M0 | 18.90 | 17.49 | 15.95 | 14.41 | 12.54 | 11.55 | 10.92 | 2.98e+01 | - | 2.09e+01 | - | 2.69E+01 |
| Sz 56 | M4 | 18.53 | 17.08 | 15.41 | 13.47 | 11.49 | 10.78 | 10.41 | 3.52e+01 | - | 2.17e+01 | - | 5.04E+01 |
| Sz 57 | M4 | 19.56 | 17.70 | 15.64 | 13.48 | 10.95 | 10.22 | 9.80 | 6.05e+01 | - | 4.12e+01 | - | 3.26e+01 |
| Sz 58 | K5 | 17.89 | 16.01 | 14.44 | 13.00 | 10.84 | 9.58 | 8.75 | 2.72e+02 | - | 2.52e+02 | - | 3.49e+02 |
| Sz 59 | M0 | 16.37 | 14.80 | 13.42 | 12.08 | 10.51 | 9.26 | 8.38 | 3.98e+02 | - | 3.07e+02 | - | 2.38e+02 |
| Sz 60a | M1 | 17.56 | 16.21 | 14.88 | 13.45 | 11.19 | 10.22 | 9.54 | 6.40e+01 | - | 4.74e+01 | - | 5.66e+01 |
| Sz 60b | M4 | 18.16 | 16.80 | 15.32 | 13.60 | 11.51 | 9.74 | 9.46 | 5.09e+01 | - | 3.97e+01 | - | - |
| Sz 61 | K4 | 16.77 | 15.13 | 13.69 | 12.38 | 9.88 | 8.76 | 7.94 | 5.94e+02 | 6.12e+02 | 5.04e+02 | 5.32e+02 | 6.50E+02 |
| Sz 62 | M2 | 16.99 | 15.55 | 14.03 | 12.56 | 10.52 | 9.65 | 9.12 | 1.35e+02 | 1.13e+02 | 9.33e+01 | 1.03e+02 | 1.16e+02 |

**NOTE.** 5 of the 20 PMS stars presented in Table 5 of Hughes & Hartigan (1992 ) paper were excluded from our analysis. Sz 47 and Sz 49 were excluded because they are heavily veiled and their spectral types are very uncertain. IRAS 12496-7650 was excluded because it is a highly embedded Herbig Be/Ae star. Sz 63 and Sz 64 were excluded because of the lack of *Spitzer* data. Even though the spectral type of Sz 55 is marked as uncertain, this star was kept on the sample because we were able to obtain a reasonable star + disk model to fit the optical, near-IR and *Spitze*r data (See figure 10).

**Table 2. Adopted extinction relations and zero points**

| Band | $\lambda$ (μm) | $A_V/A_\lambda$ ($R_V$=3.1) | $A_V/A_\lambda$ ($R_V$=5.0) | Zero point Jy |
|---|---|---|---|---|
| B | 0.44 | 1.31 | 1.20 | 4130 |
| V | 0.55 | 1.00 | 1.00 | 3781 |
| $R_C$ | 0.65 | 0.79 | 0.84 | 3080 |
| $I_C$ | 0.80 | 0.58 | 0.62 | 2550 |
| J | 1.25 | 0.26 | 0.26 | 1594 |
| H | 1.66 | 0.15 | 0.15 | 1024 |
| K | 2.2 | 0.09 | 0.09 | 667 |

NOTE: Extinction curves and optical zero points are from the Asiago database of photometric systems[6] (Fiorucci & Munari 2002) . The 2MASS zero-points are from 2MASS All Sky data release web document[7]

**Table 3. Taurus wTTs from the c2d Legacy Project**

| Star ID | SpT | V | $R_C$ | $I_C$ | J | H | K |
|---|---|---|---|---|---|---|---|
| FX Tau | M4 | 13.50 | 12.37 | 10.98 | 9.39 | 8.40 | 7.92 |
| HD 283572 | G2 | 9.05 | 8.56 | 8.07 | 7.41 | 7.01 | 6.87 |
| IW Tau | K7 | 12.51 | 11.57 | 10.51 | 9.24 | 8.48 | 8.28 |
| Lk 19 | K0 | 10.94 | 10.35 | 9.75 | 8.87 | 8.32 | 8.15 |
| LkCa 4 | K7 | 11.69 | 10.97 | 10.28 | 9.34 | 8.71 | 8.58 |
| LkCa 1 | M4 | 13.73 | 12.63 | 11.05 | 9.64 | 8.87 | 8.62 |
| LkCa 21 | M3 | 13.43 | 12.32 | 10.88 | 9.46 | 8.67 | 8.45 |
| LkCa 3 | M1 | 12.06 | 11.04 | 9.76 | 8.36 | 7.62 | 7.42 |
| LkCa 5 | M2 | 13.54 | 12.54 | 11.29 | 9.97 | 9.29 | 9.05 |
| LkCa 7 | K7 | 12.52 | 11.60 | 10.46 | 9.13 | 8.38 | 8.26 |
| NTTS 032641+2420 | K1 | 12.20 | 11.64 | 11.13 | 10.32 | 9.86 | 9.70 |
| NTTS 040234+2143 | M2 | 14.77 | 13.72 | 12.31 | 10.95 | 10.29 | 10.06 |
| NTTS 041559+1716 | K6 | 12.23 | 11.56 | 10.88 | 10.03 | 9.42 | 9.27 |
| NTTS 042417+1744 | K1 | 10.35 | 9.89 | 9.47 | 8.78 | 8.39 | 8.30 |
| NTTS 042835+1700 | K5 | 12.57 | 11.86 | 11.18 | 10.28 | 9.71 | 9.50 |
| NTTS 042916+1751 | K76 | 12.01 | 11.26 | 10.53 | 9.70 | 9.06 | 8.85 |
| NTTS 042950+1757 | K7 | 13.11 | 12.20 | 11.27 | 10.16 | 9.46 | 9.31 |
| RX J0405.3+2009 | K1 | 10.67 | 9.96 | 9.41 | 8.69 | 8.19 | 8.09 |
| RX J0409.2+1716 | M0 | 13.44 | 12.11 | 11.15 | 9.96 | 9.25 | 9.05 |
| RX J0409.8+2446 | M1 | 13.51 | 12.55 | 11.35 | 10.10 | 9.45 | 9.25 |
| RX J0412.8+1937 | K6 | 12.47 | 11.68 | 10.85 | 9.99 | 9.43 | 9.24 |
| RX J0420.3+3123 | K4 | 12.60 | 11.96 | 11.30 | 10.45 | 9.88 | 9.73 |
| RX J0432.8+1735 | M2 | 13.66 | 12.60 | 11.32 | 10.00 | 9.23 | 9.02 |
| 1RX J0438.2+202 | K2 | 12.18 | 11.52 | 10.90 | 10.07 | 9.53 | 9.36 |
| RX J0438.6+1546 | K1 | 10.89 | 10.31 | 9.73 | 8.90 | 8.36 | 8.24 |
| RX J0439.4+3332A | K5 | 11.54 | 10.79 | 10.13 | 9.18 | 8.57 | 8.42 |
| RX J0445.8+1556 | G5 | 9.29 | 8.84 | 8.41 | 7.85 | 7.46 | 7.34 |

[6] http://ulisse.pd.astro.it/Astro/ADPS/
[7] http://www.ipac.caltech.edu/2mass/releases/allsky/doc/sec6_4a.html

| | | | | | | | |
|---|---|---|---|---|---|---|---|
| RX J0452.5+1730 | K4 | 11.97 | 11.08 | 10.58 | 9.97 | 9.41 | 9.25 |
| RX J0452.8+1621 | K6 | 11.74 | 10.81 | 10.05 | 9.10 | 8.48 | 8.28 |
| RX J0457.2+1524 | K1 | 10.21 | 9.67 | 9.13 | 8.38 | 7.91 | 7.75 |
| RX J0457.5+2014 | K3 | 11.34 | 10.73 | 10.15 | 9.28 | 8.82 | 8.69 |
| RX J0458.7+2046 | K7 | 11.95 | 11.05 | 10.43 | 9.59 | 8.96 | 8.80 |
| RX J0459.7+1430 | K4 | 11.71 | 11.10 | 10.53 | 9.66 | 9.09 | 8.95 |
| UX Tau A | K2 | 11.93 | 11.11 | 10.16 | 8.62 | 7.96 | 7.55 |
| V807 Tau | K7 | 11.44 | 10.56 | 9.58 | 8.15 | 7.36 | 6.96 |
| V836 Tau | K7 | 13.99 | 12.93 | 11.74 | 9.91 | 9.08 | 8.60 |
| V927 Tau | M54 | 14.70 | 13.39 | 11.43 | 9.73 | 9.06 | 8.77 |
| V928 Tau | M0 | 14.04 | 12.77 | 11.33 | 9.54 | 8.43 | 8.11 |
| Wa Tau 1 | K0 | 10.30 | 9.76 | 9.24 | 8.42 | 7.93 | 7.80 |

## Table 4. Taurus wTTs from Strom et al. (1989)

| HBC ID | SpT | B | V | R | I | J | H | K |
|---|---|---|---|---|---|---|---|---|
| 347 | K1 | 12.95 | 12.05 | 11.52 | 11.01 | 10.32 | 9.86 | 9.70 |
| 351 | K5 | 13.38 | 12.25 | 11.55 | 10.84 | 9.80 | 9.21 | 9.07 |
| 352 | G0 | 12.71 | 11.85 | 11.34 | 10.82 | 10.08 | 9.71 | 9.58 |
| 353 | G5 | 13.25 | 12.31 | 11.75 | 11.18 | 10.45 | 10.01 | 9.86 |
| 354 | K3 | 14.90 | 13.79 | 13.10 | 12.47 | 11.79 | 11.23 | 11.09 |
| 355 | K2 | 13.59 | 12.67 | 12.13 | 11.60 | 10.81 | 10.34 | 10.21 |
| 357 | K2 | 13.96 | 12.91 | 12.28 | 11.67 | 10.84 | 10.32 | 10.16 |
| 358 | M2 | 15.99 | 14.52 | 13.40 | 11.85 | 10.27 | 9.70 | 9.46 |
| 359 | M2 | 15.07 | 14.17 | 13.09 | 11.74 | 10.37 | 9.75 | 9.53 |
| 360 | M3 | 16.54 | 14.97 | 13.72 | 12.24 | 10.80 | 10.17 | 9.97 |
| 361 | M3 | 16.61 | 15.09 | 13.92 | 12.41 | 10.94 | 10.35 | 10.10 |
| 362 | M2 | 16.04 | 14.67 | 13.60 | 12.30 | 10.95 | 10.29 | 10.06 |
| 365 | M4 | 15.22 | 13.73 | 12.52 | 11.07 | 9.64 | 8.87 | 8.62 |
| 368 | M1 | 13.61 | 12.10 | 11.01 | 9.78 | 8.36 | 7.62 | 7.42 |
| 370 | K7 | 13.96 | 12.49 | 11.54 | 10.56 | 9.25 | 8.52 | 8.32 |
| 371 | M2 | 15.06 | 13.56 | 12.51 | 11.33 | 9.98 | 9.29 | 9.05 |
| 372 | K5 | 14.37 | 13.26 | 12.60 | 11.99 | 11.18 | 10.60 | 10.46 |
| 376 | K7 | 13.41 | 12.28 | 11.59 | 10.92 | 10.03 | 9.42 | 9.27 |
| 378 | K7 | 14.81 | 13.24 | 12.24 | 11.16 | 9.50 | 8.65 | 8.42 |
| 379 | K7 | 13.94 | 12.55 | 11.63 | 10.58 | 9.13 | 8.33 | 8.26 |
| 380 | G5 | 9.87 | 9.04 | 8.34 | 7.83 | 7.41 | 7.01 | 6.87 |
| 385 | M1 | 14.50 | 13.04 | 12.06 | 11.09 | 9.78 | 8.89 | 8.35 |
| 388 | K1 | 11.13 | 10.34 | 9.88 | 9.45 | 8.78 | 8.39 | 8.30 |
| 392 | K5 | 13.71 | 12.53 | 11.81 | 11.15 | 10.28 | 9.71 | 9.50 |
| 397 | K7 | 13.29 | 12.06 | 11.31 | 10.61 | 9.70 | 9.06 | 8.85 |
| 399 | K7 | 13.58 | 12.18 | 11.29 | 10.34 | 9.17 | 8.49 | 8.23 |
| 400 | K7 | 13.51 | 12.11 | 11.23 | 10.35 | 9.07 | 8.43 | 8.25 |
| 403 | K7 | 14.72 | 13.22 | 12.28 | 11.37 | 10.16 | 9.46 | 9.31 |
| 407 | F8 | 13.70 | 12.67 | 12.04 | 11.43 | 10.58 | 10.08 | 9.90 |
| 408 | K0 | 11.33 | 10.37 | 9.80 | 9.27 | 8.42 | 7.93 | 7.80 |
| 415 | G0 | 12.46 | 11.07 | 10.21 | 9.36 | 8.10 | 7.50 | 7.23 |
| 419 | K5 | 13.41 | 12.09 | 11.31 | 10.57 | 9.42 | 8.60 | 8.16 |
| 420 | K7 | 14.04 | 12.51 | 11.51 | 10.51 | 9.24 | 8.48 | 8.28 |
| 426 | K0 | 11.87 | 10.85 | 10.25 | 9.68 | 8.87 | 8.32 | 8.15 |
| 427 | K7 | 12.88 | 11.60 | 10.81 | 10.05 | 8.96 | 8.32 | 8.13 |
| 429 | K7 | 14.66 | 13.13 | 12.19 | 11.21 | 9.91 | 9.08 | 8.60 |

**Table 5. Taurus cTTs from Strom et al. (1989)**

| HBC ID | SpT | B | V | $R_C$ | $I_C$ | J | H | K | $J_x$ SED | $r_J$ Spectra | $\sigma r_J$ (*) |
|---|---|---|---|---|---|---|---|---|---|---|---|
| 23 | M0 | 15.01 | 14.30 | 13.50 | 12.40 | 10.32 | 9.39 | 8.71 | 1.01 | 1.02 | 0.52 |
| 25 | K3 | 13.59 | 12.36 | 11.42 | 10.60 | 9.00 | 7.87 | 6.86 | -0.01 | 1.2 | LL |
| 26 | M2 | 15.46 | 13.91 | 12.78 | 11.39 | 9.96 | 9.15 | 8.84 | 0.04 | 0.80 | LL |
| 27 | M2 | 15.24 | 13.67 | 12.57 | 11.27 | 9.87 | 9.05 | 8.81 | - | - | - |
| 28 | M0 | 15.03 | 14.11 | 13.09 | 11.55 | 9.56 | 8.62 | 7.97 | 0.05 | 0.43 | 0.21 |
| 30 | M2 | 15.03 | 14.11 | 13.09 | 11.55 | 9.56 | 8.62 | 7.97 | 0.04 | 0.99 | 0.21 |
| 32 | K7 | 13.07 | 12.06 | 11.23 | 10.39 | 9.30 | 8.42 | 8.05 | 0.45 | 0.52 | 0.14 |
| 33 | M2 | 14.31 | 12.95 | 11.87 | 10.66 | 9.15 | 8.26 | 7.71 | 0.32 | 0.41 | 0.15 |
| 34 | K1 | 11.86 | 10.92 | 10.28 | 9.63 | 8.00 | 6.78 | 5.74 | 3.15 | 0.80 | LL |
| 35 | K0 | 11.07 | 9.89 | 9.11 | 8.45 | 7.15 | 6.18 | 5.40 | 0.22 | 0.78 | 0.15 |
| 36 | M1 | 13.17 | 12.08 | 11.08 | 9.94 | 8.17 | 7.25 | 6.73 | - | - | - |
| 37 | K7 | 13.09 | 12.27 | 11.33 | 10.48 | 8.96 | 7.81 | 6.73 | 1.25 | 2.50 | LL |
| 38 | M0 | 15.32 | 13.92 | 12.80 | 11.53 | 9.77 | 8.82 | 8.18 | - | - | - |
| 41 | M0 | 15.08 | 13.53 | 12.43 | 11.27 | 9.66 | 8.64 | 8.00 | 0.52 | 0.12 | 0.12 |
| 44 | M1 | 15.54 | 13.90 | 12.71 | 11.25 | 9.55 | 8.57 | 8.06 | 0.15 | 0.35 | 0.16 |
| 45 | K7 | 13.77 | 12.45 | 11.47 | 10.48 | 8.88 | 7.82 | 7.03 | 0.82 | 0.72 | 0.20 |
| 46 | M4 | 15.76 | 14.28 | 13.04 | 11.29 | 9.52 | 8.78 | 8.54 | 0.10 | 0.60 | 0.22 |
| 48 | M1 | 17.48 | 15.75 | 14.29 | 12.73 | 10.33 | 9.08 | 8.37 | 0.47 | 0.43 | 0.21 |
| 49 | M0 | 15.92 | 14.55 | 13.47 | 12.36 | 10.21 | 8.67 | 7.10 | 0.47 | 2.50 | LL |
| 50 | M3 | 16.37 | 14.94 | 13.59 | 11.81 | 9.34 | 8.16 | 7.17 | 0.58 | 0.8 | LL |
| 51 | K7 | 15.20 | 13.65 | 12.46 | 10.99 | 8.98 | 8.81 | 8.52 | - | - | - |
| 54 | K7 | 13.75 | 12.35 | 11.36 | 10.38 | 8.79 | 7.85 | 7.25 | 0.61 | 0.38 | 0.14 |
| 55 | M2 | 14.48 | 12.95 | 11.90 | 10.61 | 9.30 | 8.38 | 7.86 | 0.20 | 0.11 | 0.18 |
| 56 | K6 | 14.27 | 13.09 | 12.08 | 11.05 | 9.16 | 8.22 | 7.55 | 0.50 | 0.40 | 0.17 |
| 57 | K7 | 13.64 | 12.37 | 11.45 | 10.53 | 9.00 | 8.01 | 7.32 | 0.54 | 0.72 | 0.22 |
| 58 | K7 | 14.19 | 13.05 | 12.11 | 11.15 | 9.69 | 8.77 | 8.12 | 0.51 | 0.7 | LL |
| 61 | K7 | 14.72 | 13.37 | 12.32 | 11.22 | 9.48 | 8.43 | 7.79 | - | - | - |
| 62 | M1 | 15.42 | 13.99 | 12.99 | 11.85 | 10.44 | 9.76 | 9.52 | - | - | - |
| 63 | K7 | 14.23 | 12.96 | 12.05 | 11.01 | 9.67 | 8.76 | 8.29 | 0.42 | 0.21 | 0.08 |
| 65 | M0 | 13.72 | 12.33 | 11.40 | 10.44 | 9.16 | 8.37 | 8.03 | 0.21 | 0.08 | 0.10 |
| 66 | K3 | 14.83 | 13.16 | 12.07 | 11.00 | 9.34 | 8.24 | 7.47 | -0.10 | 1.06 | 0.36 |
| 67 | K7 | 15.70 | 14.30 | 13.11 | 11.79 | 9.64 | 8.44 | 7.49 | 0.90 | 1.07 | 0.28 |
| 68 | M0 | 15.26 | 13.75 | 12.71 | 11.53 | 9.85 | 9.10 | 8.79 | 0.37 | -0.04 | 0.09 |
| 69 | K7 | 16.51 | 14.74 | 13.32 | 11.96 | 9.58 | 8.40 | 7.94 | - | - | - |
| 70 | M1 | 15.65 | 14.31 | 13.20 | 12.05 | 10.99 | 9.69 | 8.76 | - | - | - |
| 71 | M0 | 16.35 | 14.89 | 13.62 | 12.30 | 10.71 | 9.77 | 9.33 | - | - | - |
| 72 | M0 | 15.14 | 13.55 | 12.41 | 11.25 | 9.56 | 8.63 | 8.08 | 0.91 | 0.15 | 0.19 |
| 73 | K6 | 14.98 | 13.42 | 12.28 | 11.09 | 9.24 | 7.99 | 7.31 | - | - | - |
| 74 | K6 | 12.23 | 11.57 | 10.91 | 10.25 | 8.54 | 7.46 | 6.45 | 0.24 | 2.50 | LL |
| 75 | K6 | 12.78 | 11.90 | 11.16 | 10.46 | 9.33 | 8.59 | 8.16 | 0.32 | 0.57 | 0.19 |
| 76 | K7 | 13.61 | 12.37 | 11.39 | 10.46 | 8.91 | 7.89 | 7.05 | 0.24 | 0.84 | 0.20 |
| 77 | K3 | 13.22 | 12.03 | 11.22 | 10.50 | 9.34 | 8.60 | 8.28 | - | - | - |
| 80 | K1 | 11.84 | 11.09 | 10.42 | 9.68 | 8.54 | 7.62 | 6.87 | 0.23 | 0.5 | LL |
| 367 | K3 | 12.02 | 10.65 | 9.80 | 8.94 | 7.63 | 6.83 | 6.48 | 0.12 | 1.37 | 0.40 |

NOTE: (1) LL stands for Lower Limit.

**Table 6. Statistics on the near-IR colors of T Tauris stars.**

| **Band** | **wTTs mean excess** | **wTTs median excess** | **wTTs std dev $\equiv 1\sigma$** | **cTTs mean excess** | **cTTs median excess** | **cTTs excess > 1 $\sigma$** | **cTTs excess > 2 $\sigma$** | **cTTs excess > 3 $\sigma$** |
|---|---|---|---|---|---|---|---|---|
| J | 0.07 | 0.06 | 0.14 | 0.35 | 0.28 | 63% | 48% | 32% |
| H | 0.06 | 0.03 | 0.18 | 0.69 | 0.54 | 78% | 64% | 52% |
| K | 0.10 | 0.06 | 0.24 | 1.45 | 1.10 | 79% | 74% | 67% |

**Table 7.**

| Band | Veiling (accretion shock) | Veiling (inner rim) | $\Delta$m ($r_V$=0.60) ($r_J$ =0.35) |
|---|---|---|---|
| $R_C$ | 0.50 $r_V$ | 0.05 $r_J$ | 0.28 |
| $I_C$ | 0.23 $r_V$ | 0.15 $r_J$ | 0.19 |

**Table 8.**

| Band | $\Delta$A (mag) | $\Delta(I_C-m)_{Av}$ (mag) | $\Delta$excess [(1)] (mag) |
|---|---|---|---|
| B | 0.56 | -0.31 | -0.50 |
| V | 0.43 | -0.18 | -0.37 |
| $R_C$ | 0.34 | -0.09 | -0.28 |
| $I_C$ | 0.25 | 0.00 | -0.19 |
| J | 0.11 | 0.14 | -0.05 |
| H | 0.06 | 0.19 | 0.00 |
| K | 0.04 | 0.21 | +0.02 |

NOTE: (1) $\Delta$excess $= \Delta(I_C-m)_{Av} - \Delta m_{Ic,}$ where $\Delta m_{Ic}$=0.19

**Table 9. Parameters of the models**

| Star ID | $L_*$ ($L_\odot$) | $T_{eff}$ (K) | $M_*$ ($M_\odot$) | $R_{rimA}$ (AU) | $\Omega$ [(1)] | $R_{rimB}$ (AU) |
|---|---|---|---|---|---|---|
| Sz 47 | 0.19 | 3470 | 0.35 | 0.04 | 4.1 | 0.07 |
| Sz 51 | 0.35 | 3850 | 0.60 | 0.05 | 5.8 | 0.10 |
| Sz 53 | 0.26 | 3720 | 0.50 | 0.04 | 3.5 | 0.07 |
| Sz 55 | 0.18 | 3850 | 0.60 | 0.04 | 1 | 0.04 |
| Sz 56 | 0.23 | 3370 | 0.30 | 0.04 | 1 | 0.04 |
| Sz 57 | 0.40 | 3370 | 0.30 | 0.05 | 1 | 0.05 |
| Sz 58 | 0.51 | 4350 | 0.60 | 0.05 | 6.0 | 0.10 |
| Sz 59 | 0.41 | 3850 | 0.40 | 0.06 | 6.7 | 0.13 |
| Sz 61 | 1.66 | 4590 | 1.40 | 0.11 | 4.5 | 0.20 |
| Sz 62 | 0.42 | 3580 | 0.40 | 0.05 | 3.6 | 0.08 |

NOTE: (1) rim area scaling factor

**Figure 1.** Optical and IR SEDs of 6 Chamaeleon II cTTs with the stellar photosphere normalized to the J-band (left column) and to the $I_C$ band (right column). In the first case, the BVRI fluxes are significantly overestimated. Normalizing the SED to the $I_C$-band leads to a better fit of the optical bands and reveals significant J and H-band excess for many of the sources. The open squares represent observed fluxes while filled circles denote extinction corrected fluxes. The solid lines indicate the expected stellar photospheres. Photometric uncertainties (usually around 3% in the optical and the near-IR, and 5% in IRAC wavelengths) are small compared to other sources of error, which include errors in the spectral types, the adopted colors, and extinction corrections. We try to quantify these errors in sections 2.2 and 2.3.

**Figure 2.** SEDs of 6 Taurus wTTs illustrate the very good agreement between expected and extinction-corrected observed fluxes for stars of different spectral types. The open squares represent observed fluxes while filled circles denote extinction corrected fluxes. The solid line indicates the expected stellar photosphere (calculated as described in section 2.1. i.e, based on expected broad band colors normalized to the $I_C$ band) and is *not* a fit to the extinction-corrected data points. The excellent agreement between the expected and extinction-corrected fluxes gives us confidence in the stellar intrinsic colors and extinction corrections that we use.

**Figure 3.** Histograms of the excess at the J-band (a and b), H-band (c), and K-band (d) for wTTs (dotted lines) and cTTs (solid lines). The excesses shown for cTTs have been corrected for the median excesses found for wTTs (third column in Table 6). The distributions shown are consistent with wTTs having no near-IR excess. Significant excess is seen in all three 2MASS bands for cTTs.

**Figure 4.** Extinction (Av) vs. J-band excess (left panel) and H-band excess (right panel). No significant correlation is seen in the figures. A shallower extinction curve with $R_V$=5.0 can only account for the IR excess for the objects left to the shown solid lines.

**Figure 5.** J-band excess calculated using our SED fitting vs. the J-band excess derived spectroscopically by FE99. The data points cluster in the upper-right quadrant of the figure; therefore, both methods show clear evidence of J-band excess for cTTs as a group.

**Figure 6.** J-band excess calculated using our SED fitting vs. $r_V$ from Gullbring et al. (1998). No significant correlation is seen with this small data set, but clearly, $r_J >> 0.1\ r_V$, instead of $r_J \sim 0.1\ r_V$, as would be expected if both originated directly at the accretion shock. The $r_J \sim 0.1\ r_V$ relation is represented by the solid line.

**Figure 7**. K-band excess vs. J-band excess (left panel) and H-band excess (right panel). K-band excess correlates strongly with both, J and H-band excesses. The 1 $\sigma$ error bars shown at the upper left corner of the figures correspond to the standard deviation of wTTs listed in Table 6. The Spearman's ranks of these correlations are 0.65 and 0.92 with probabilities of being drawn from a random distribution of $1.5\times10^{-8}$ and $6.3\times10^{-26}$

respectively. These robust correlations suggest that the J, H, and K-band excesses have a common source.

**Figure 8.** K-band excess *flux* vs. J-band excess *flux* (left panel) and H-band excess *flux* (right panel) in units of the expected photosphere at 2.2 µm. The flux ratios shown are consistent with black body emission at a relatively narrow range of temperatures, T ~ $1750 \pm 250$ K. The lines shown correspond to black bodies at 1500, 1750, and 2000 K. The 1 $\sigma$ error bars shown at the upper left corner of the figures correspond to the standard deviation of wTTs listed in Table 6. For reasons discussed in section 2.3, the actual error bars are probably larger (see text).

**Figure 9.** IRAC-1 excess *flux* vs. IRAC-3 excess *flux,* in units of the expected photosphere at 5.8 µm. The flux ratios shown are consistent with black body emission at T ~ $1400 \pm 200$ K. The lines shown correspond to black bodies at 1200, 400, and 1700 K.

**Figure 10.** Disk models for 10 chamaeleon II cTTs. The solid blue line (Total SED A) corresponds to the total SED when the inner rim is irradiated only by the *photosphere* of the central star (rim A). The solid red line (Total SED B) corresponds to the total SED when the emission from the inner rim is scaled by the factor $\Omega$ listed in Table 9 (rim B). According to the models, the near and mid-IR SED of the cTTs is largely dominated by the emission from the inner rim at the dust sublimation temperature T ~ $1750 \pm 250$ K. Also, in most cases, the area of the inner rim is larger than expected for a rim in thermal equilibrium with the stellar radiation field alone. Thus, an additional source of energy is needed. We argue that as proposed by D'Alessio et al. (2003), the UV radiation from the accretion shock significantly affects the sizes of the inner holes in disks around cTTs increasing the area of the inner rim.

**Figure 11.** Accretion luminosity ($L_{acc}$) vs. K-band excess. The plot shows that K-band excess correlates well with accretion shock luminosity. The Spearman's rank of the correlation is 0.81 with a probability of being drawn from a random distribution of $1.01 \times 10^{-6}$ . The figure also shows that, for some cTTs, the accretion luminosity can dominate the stellar luminosity.

**Figure 12.** Stellar ages derived from J and $I_C$-band luminosities for 3 different sets of evolutionary tracks: D'Antona et al (1998) (a) , Siess et al (2000) (b) and Beraffe et al. (1998) (c). In all cases, cTTs appear systematically younger when the luminosities are derived from the J-band with respect to ages obtained from the $I_C$-band luminosities. No systematic effect is seen on the derived ages of wTTs.

**Figure 13.** Age distribution of cTTs and wTTs when the ages are estimated from J-band luminosities (left panel) and when the ages are derived using $I_C$-band luminosities (right panel). The ages correspond to the models by Siess et al (2000).We suggest that, when J-band luminosities are used, the J-band excess artificially narrows the apparent age distribution of cTTs, and displaces its mean age to a younger value.

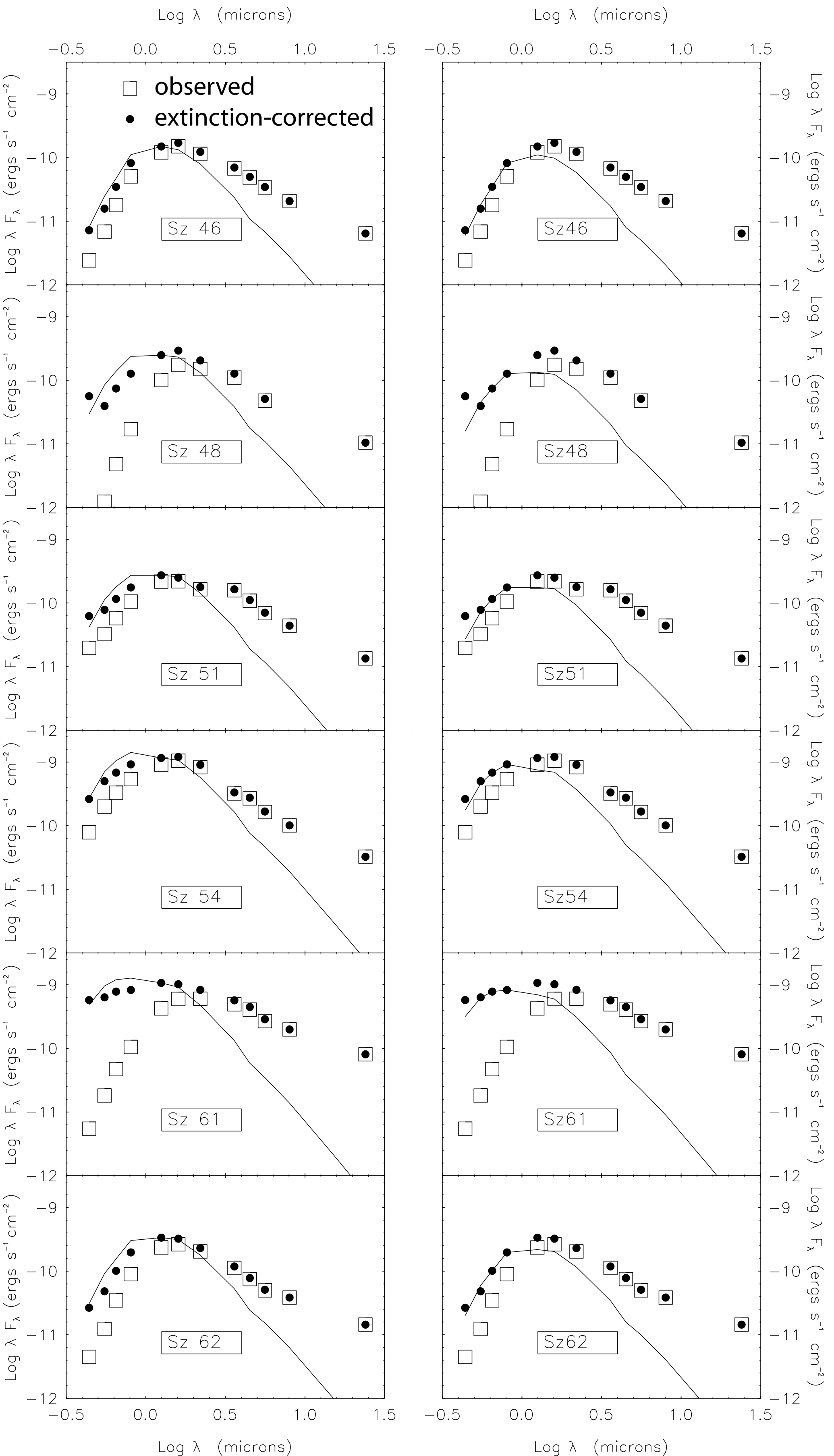

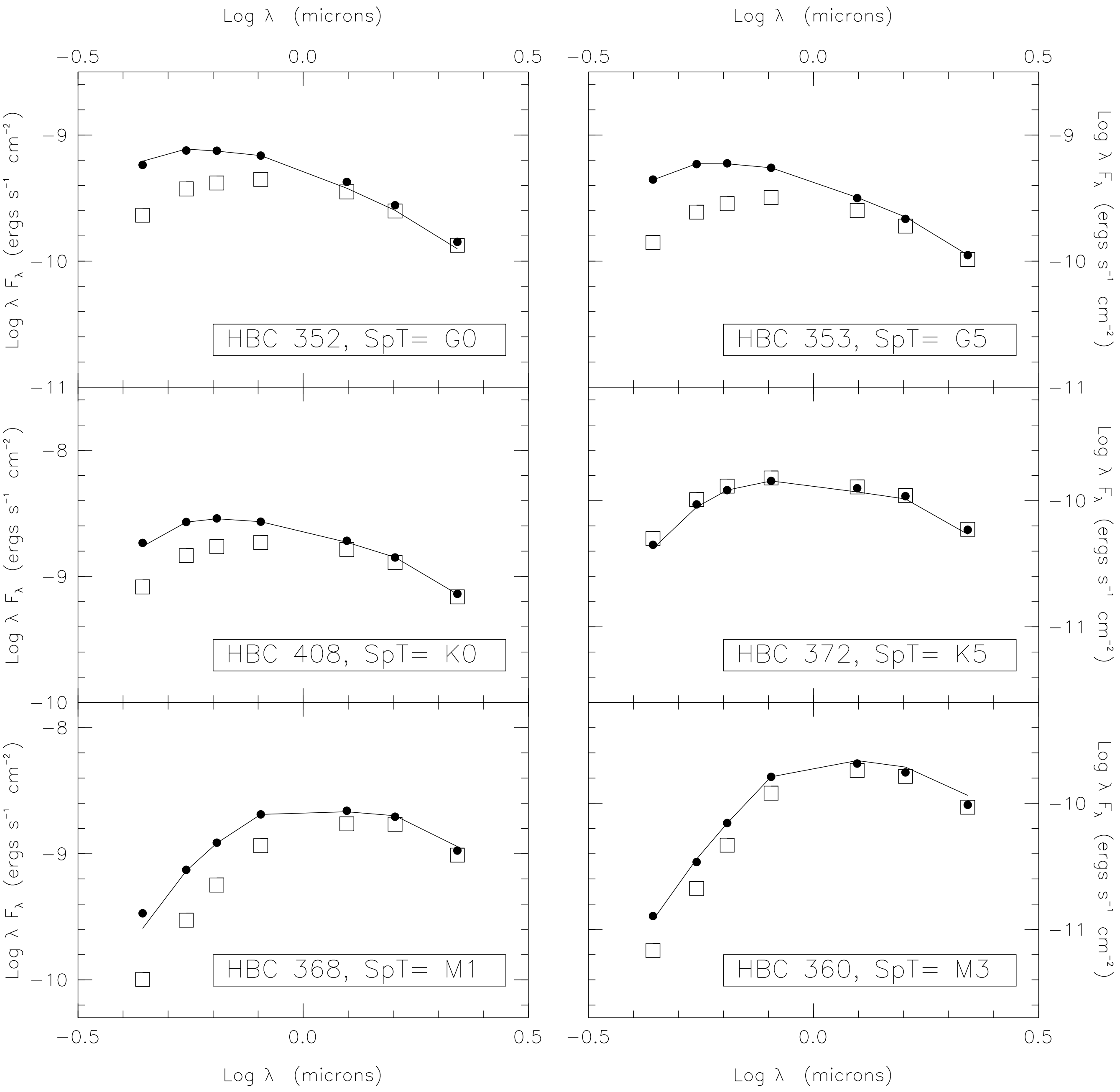
Log λ (microns)
Log λ F_λ (ergs s⁻¹ cm⁻²)
HBC 352, SpT= G0
HBC 353, SpT= G5
HBC 408, SpT= K0
HBC 372, SpT= K5
HBC 368, SpT= M1
HBC 360, SpT= M3

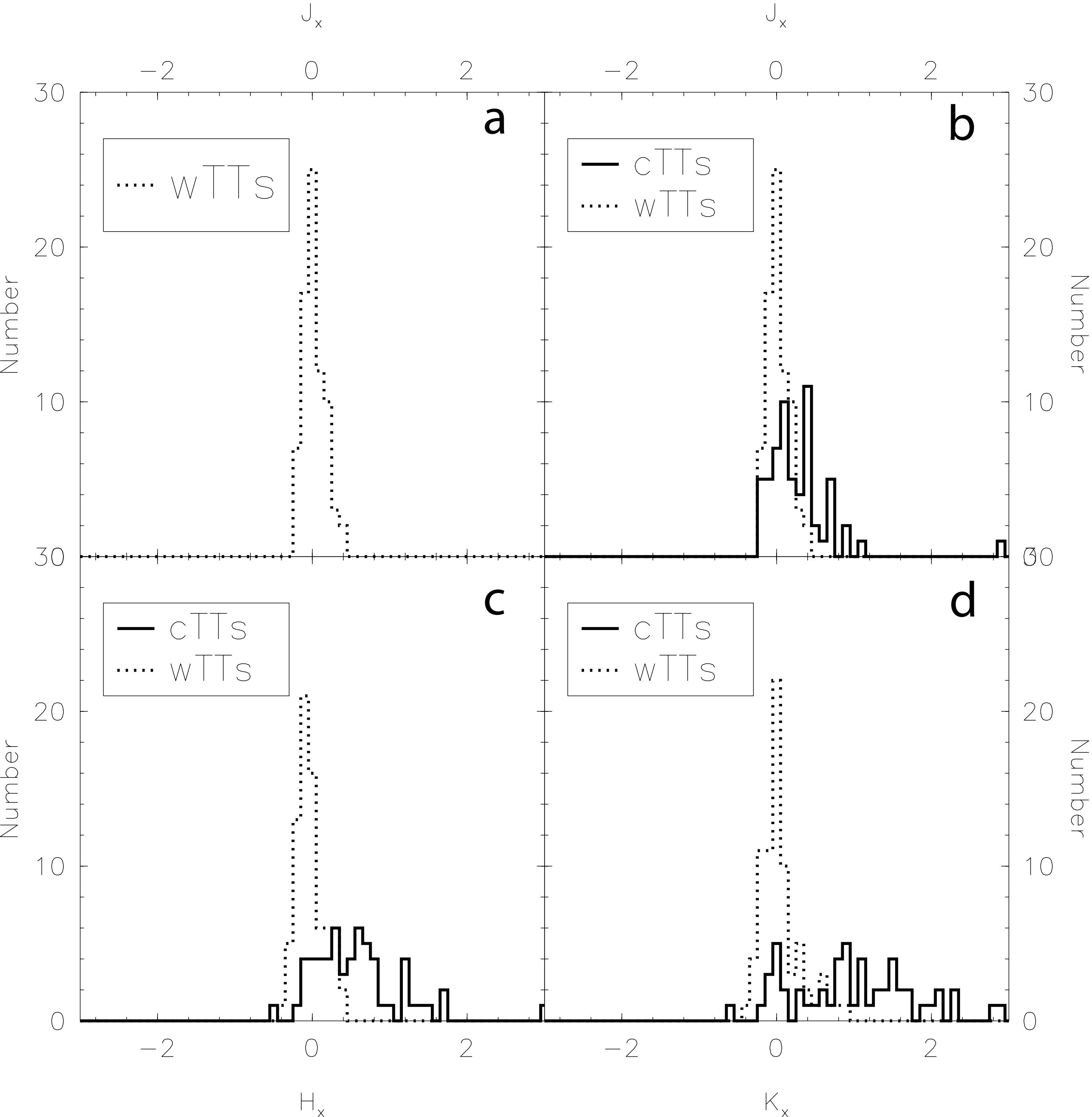

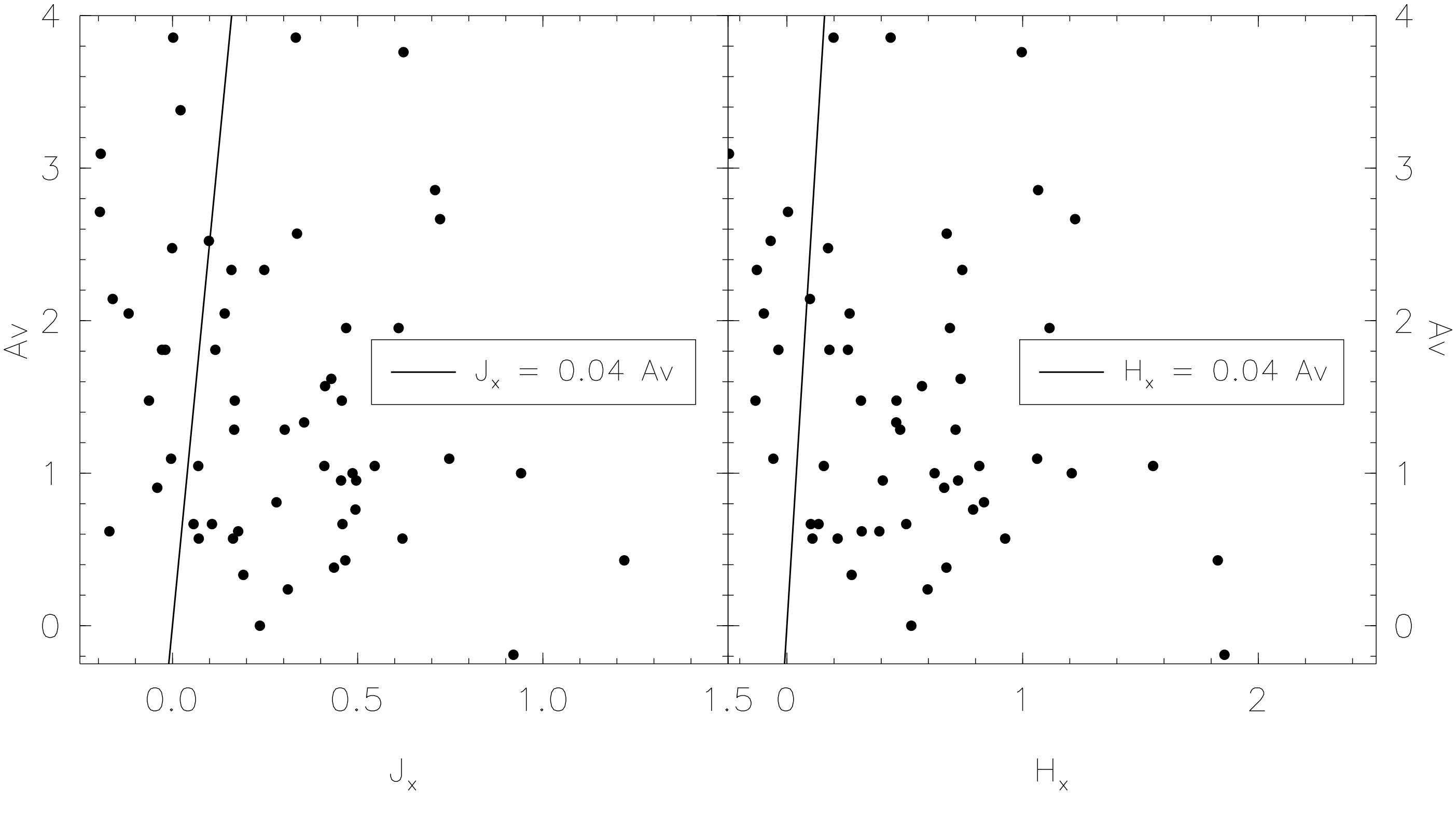

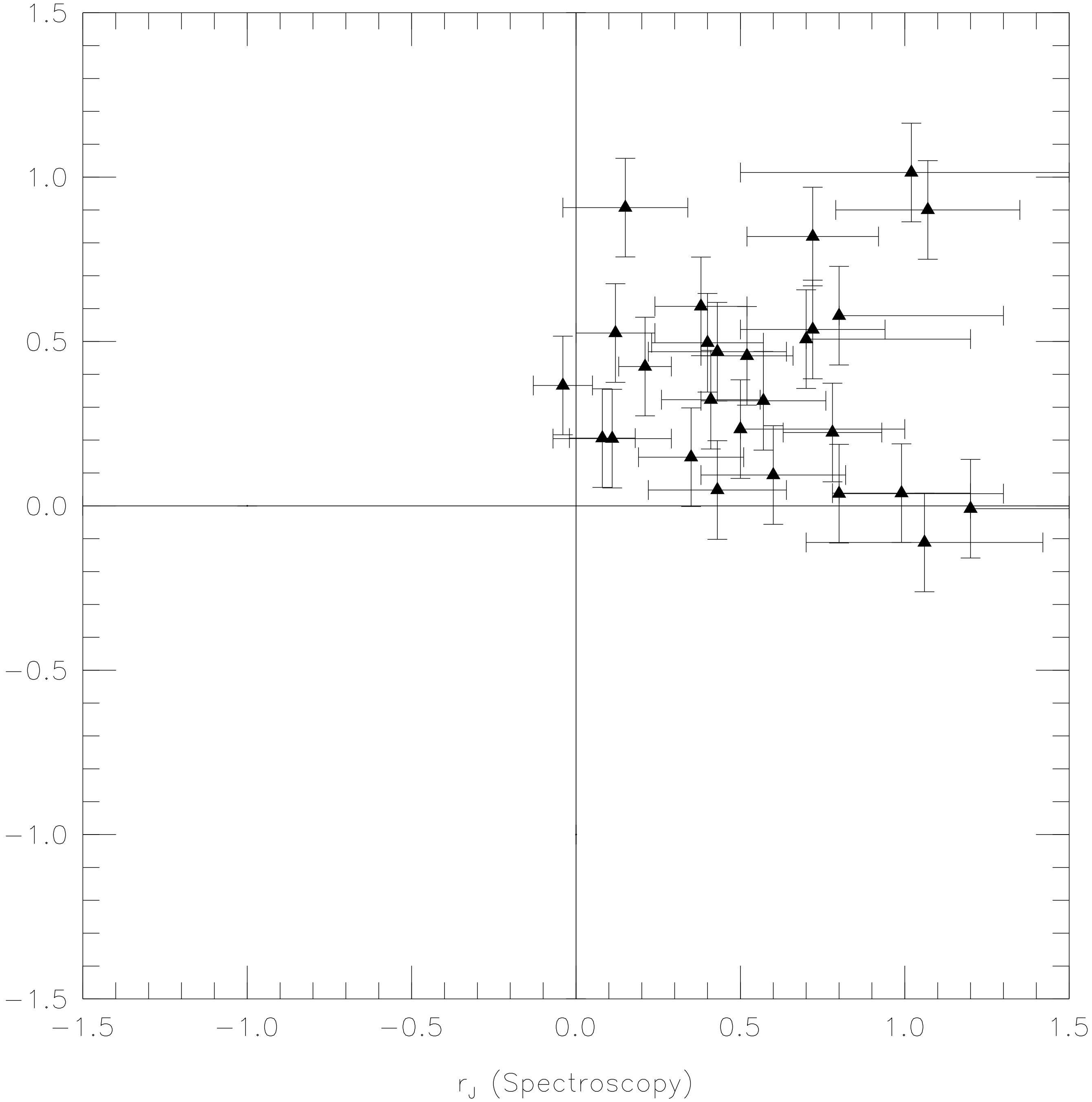

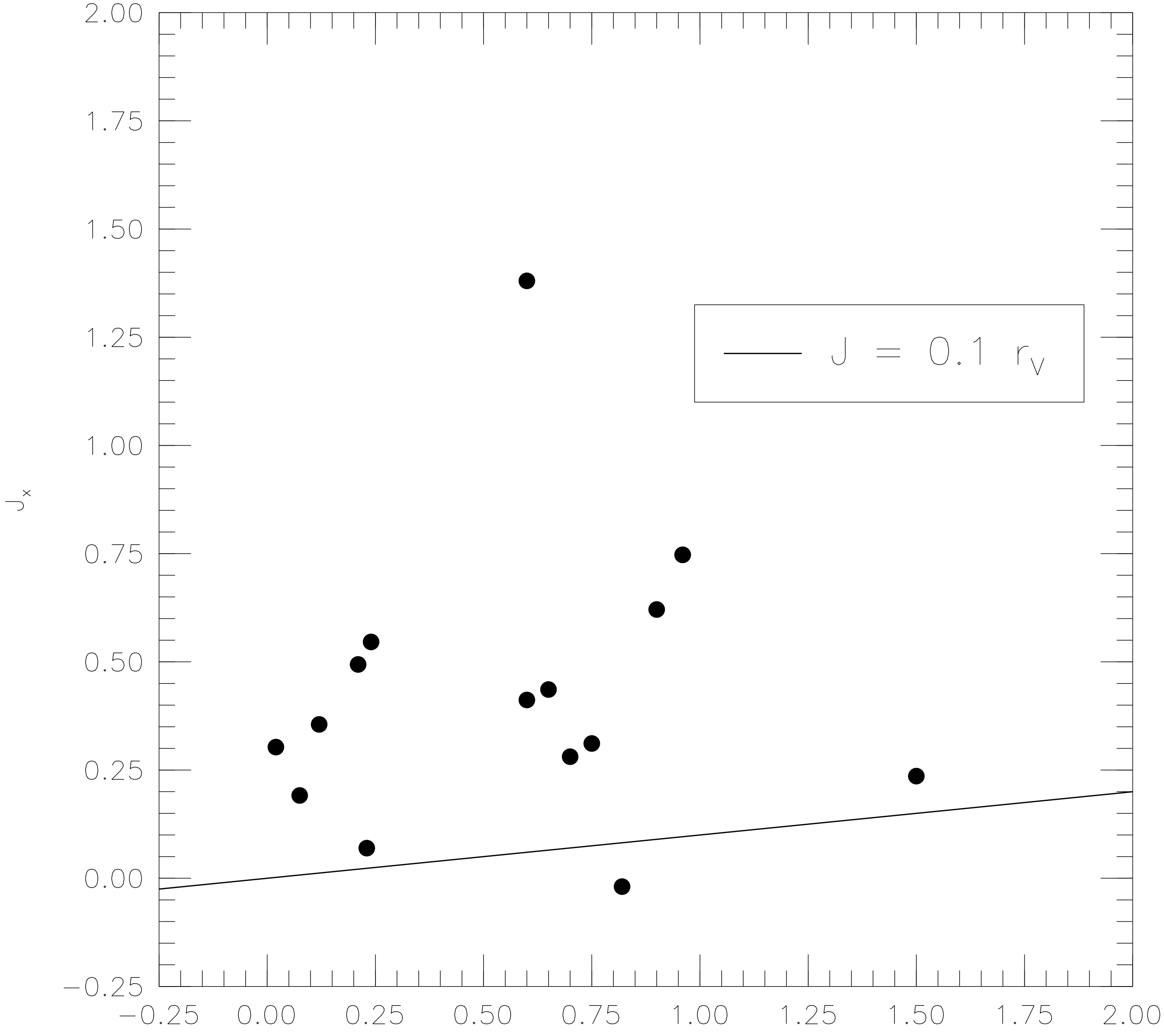

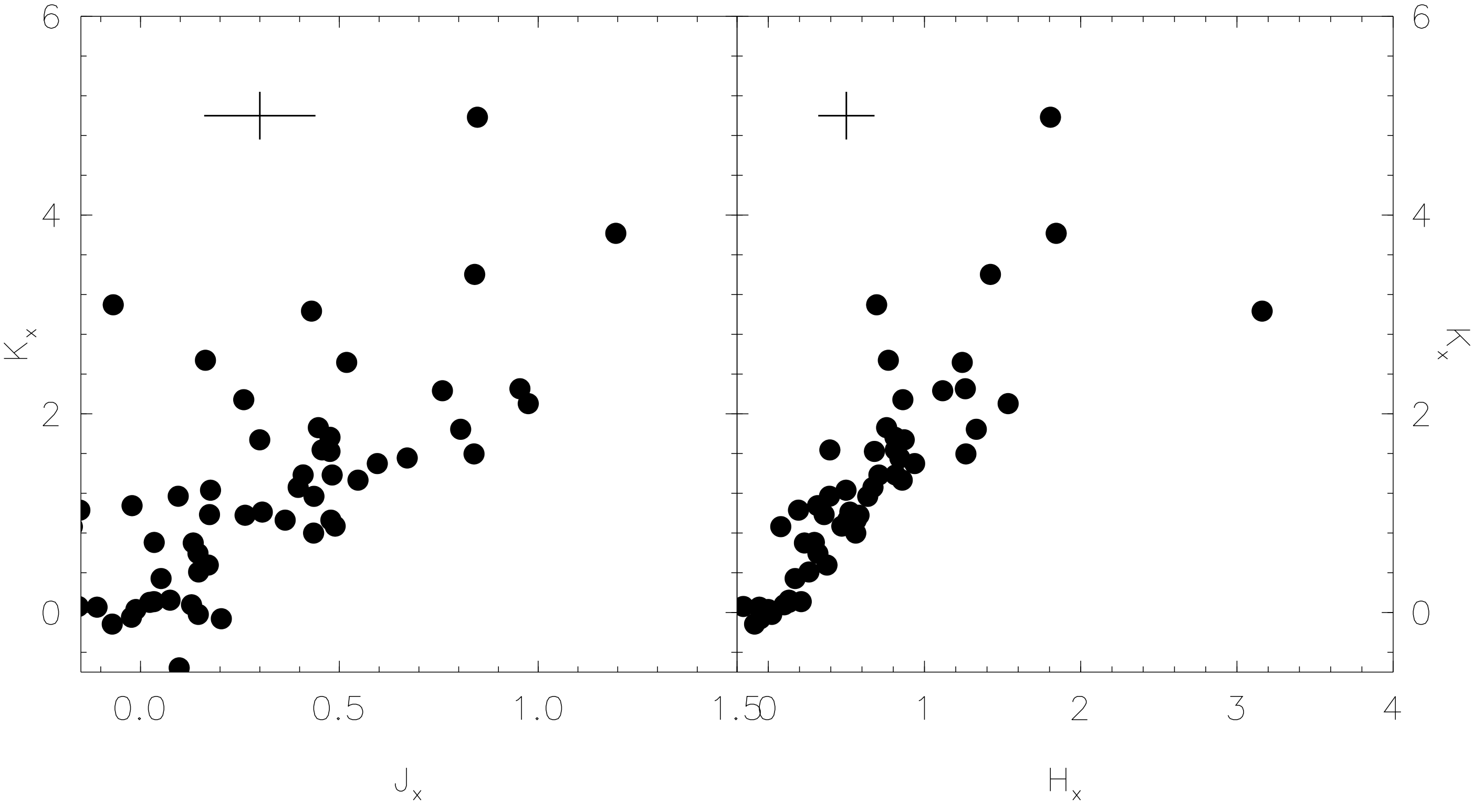

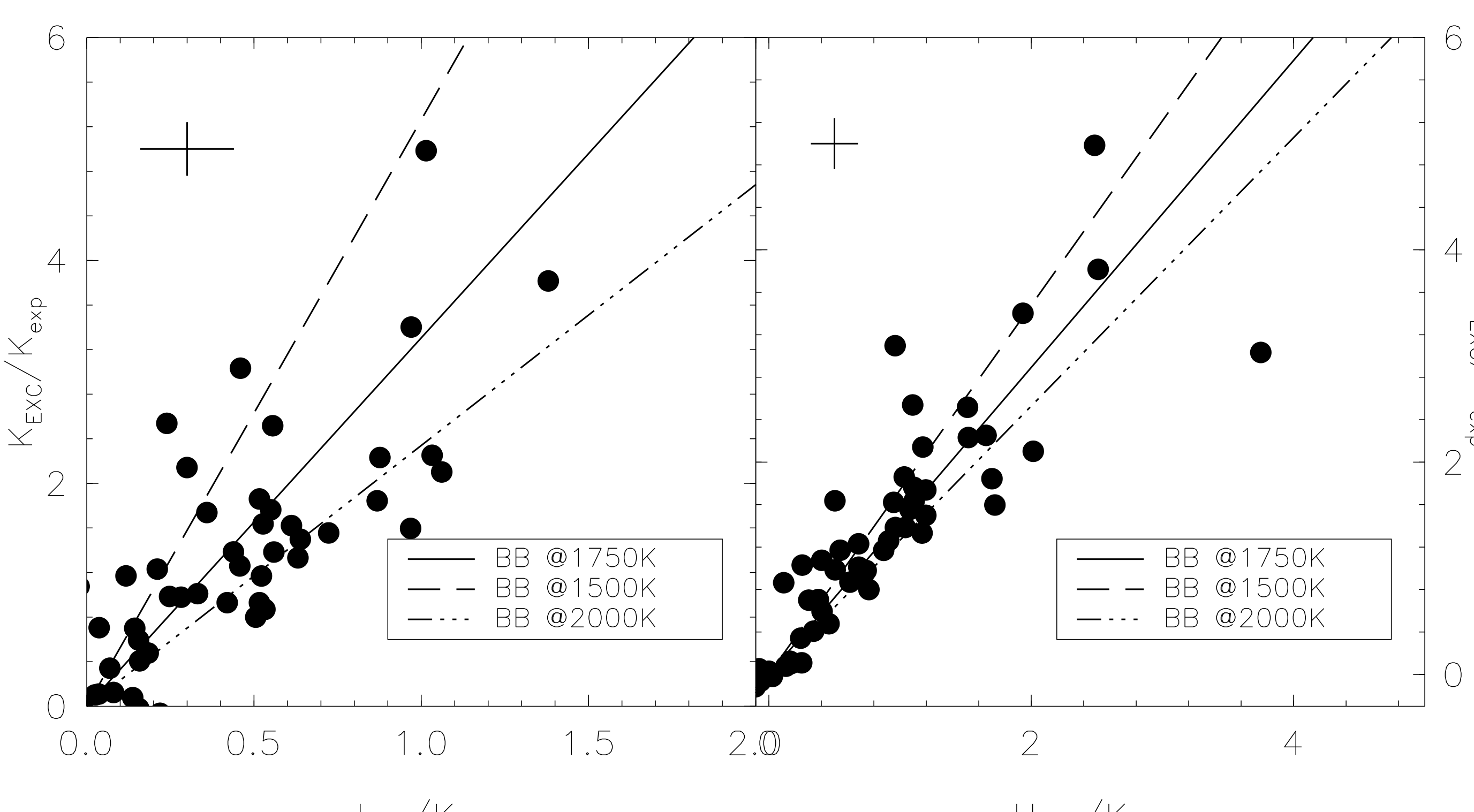

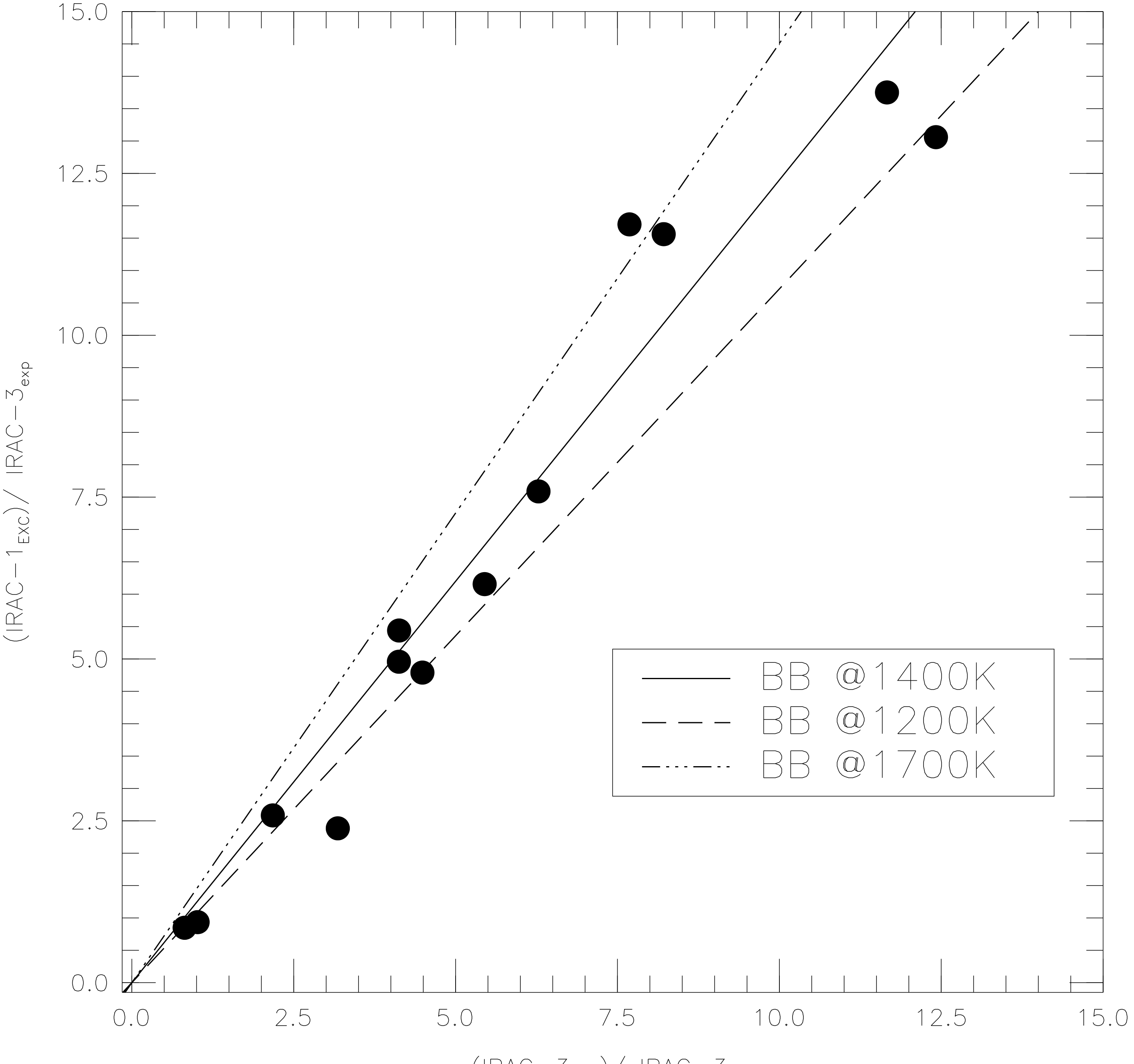

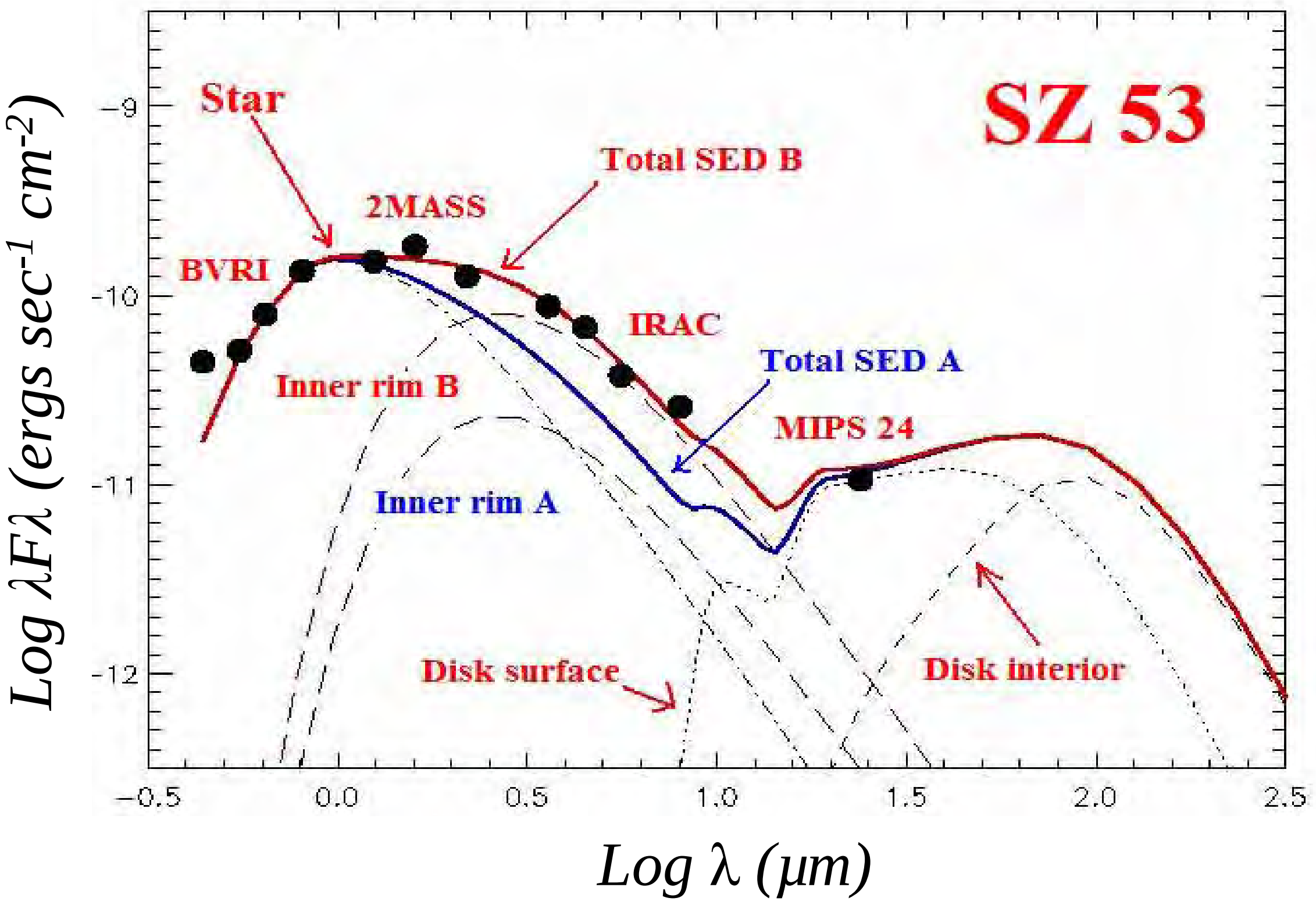

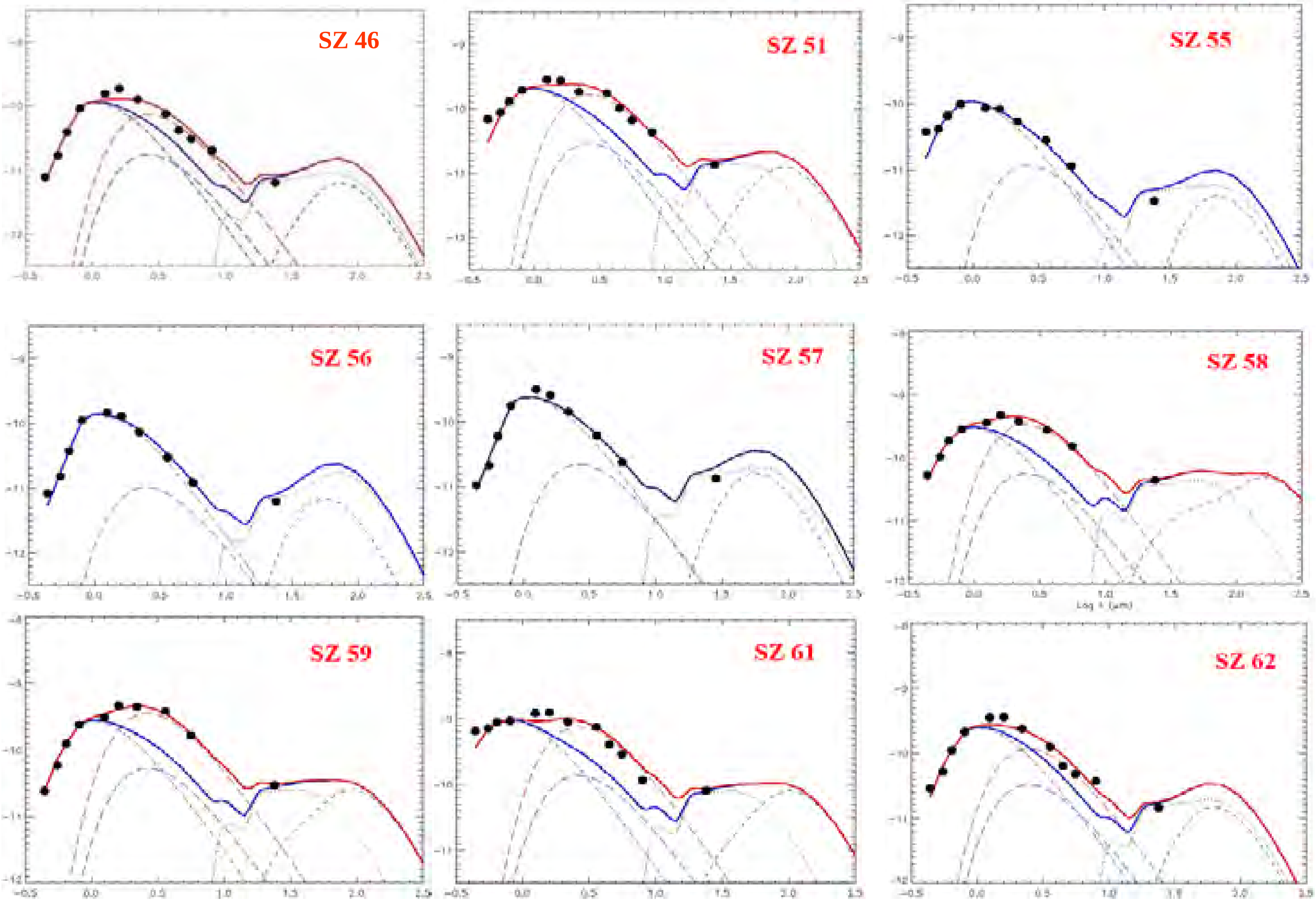

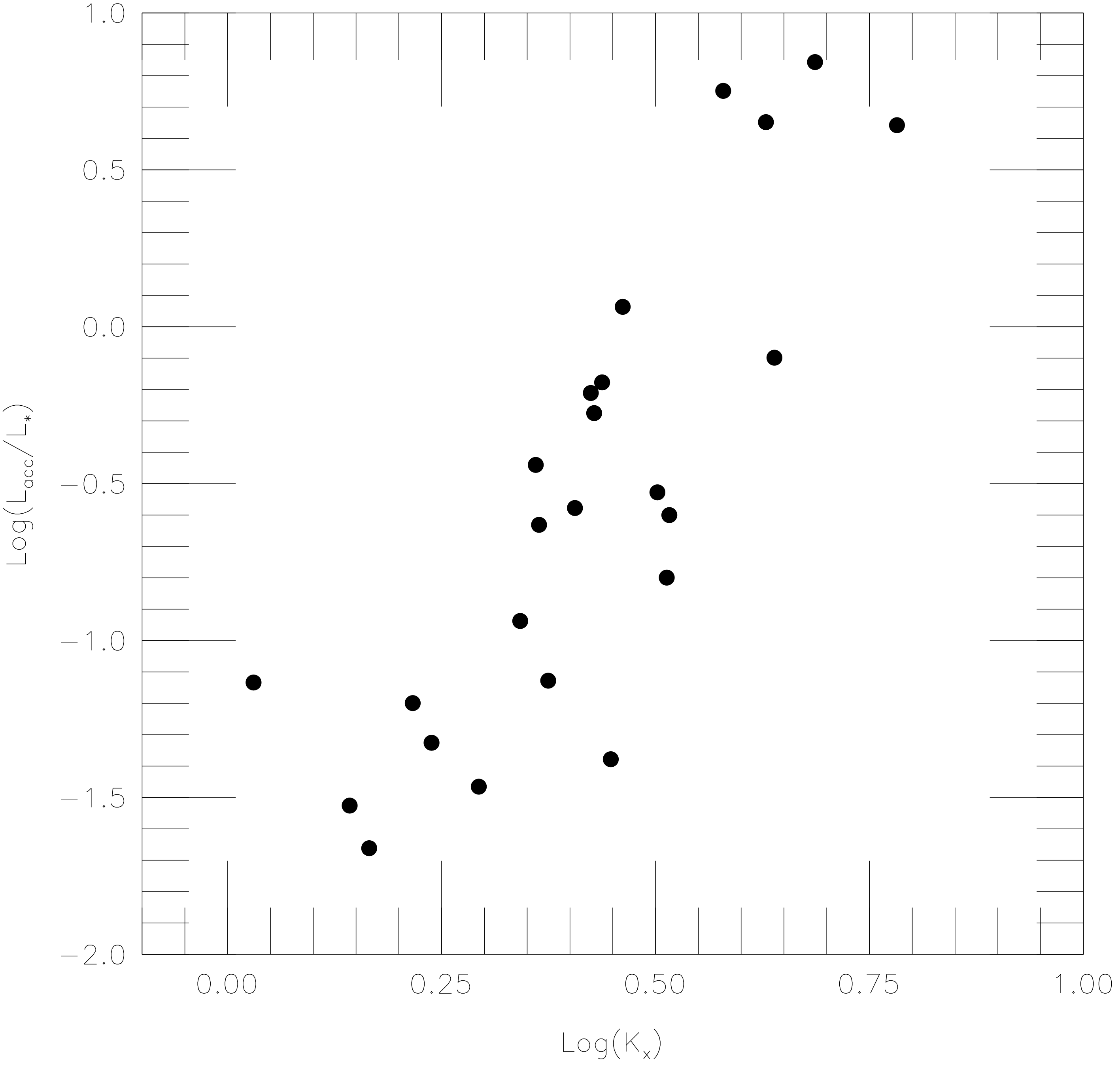

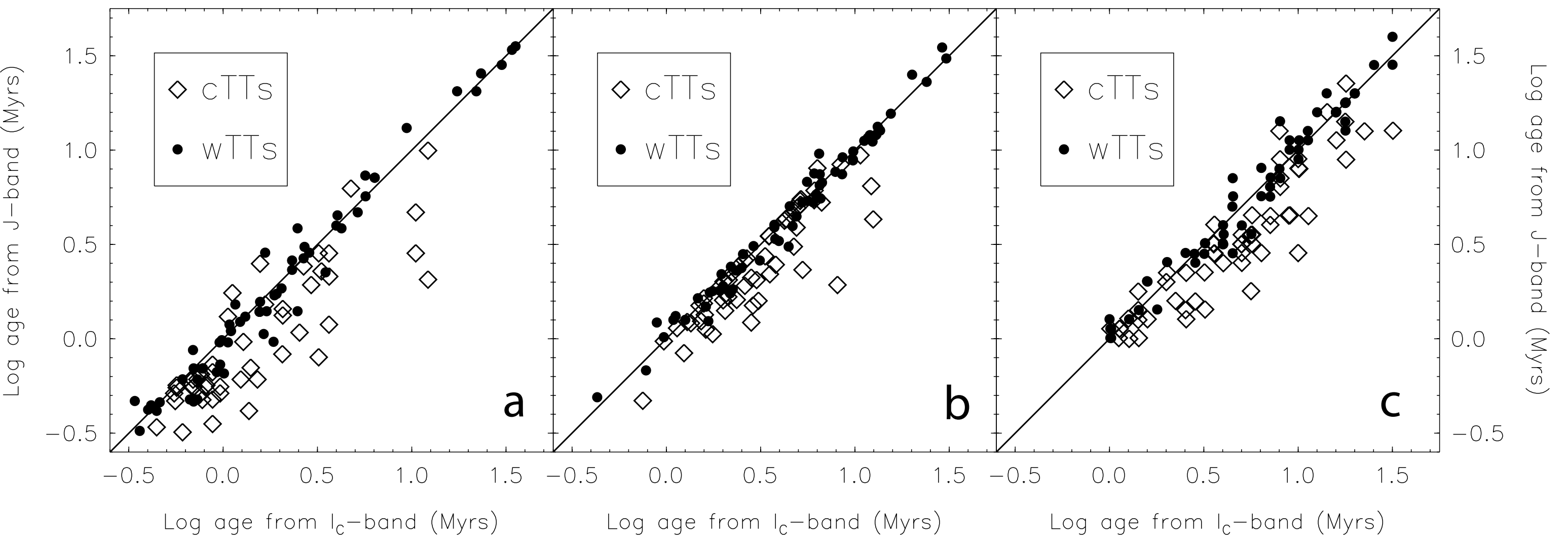

cTTs
wTTs
a
b
c
Log age from J–band (Myrs)
Log age from $I_c$–band (Myrs)

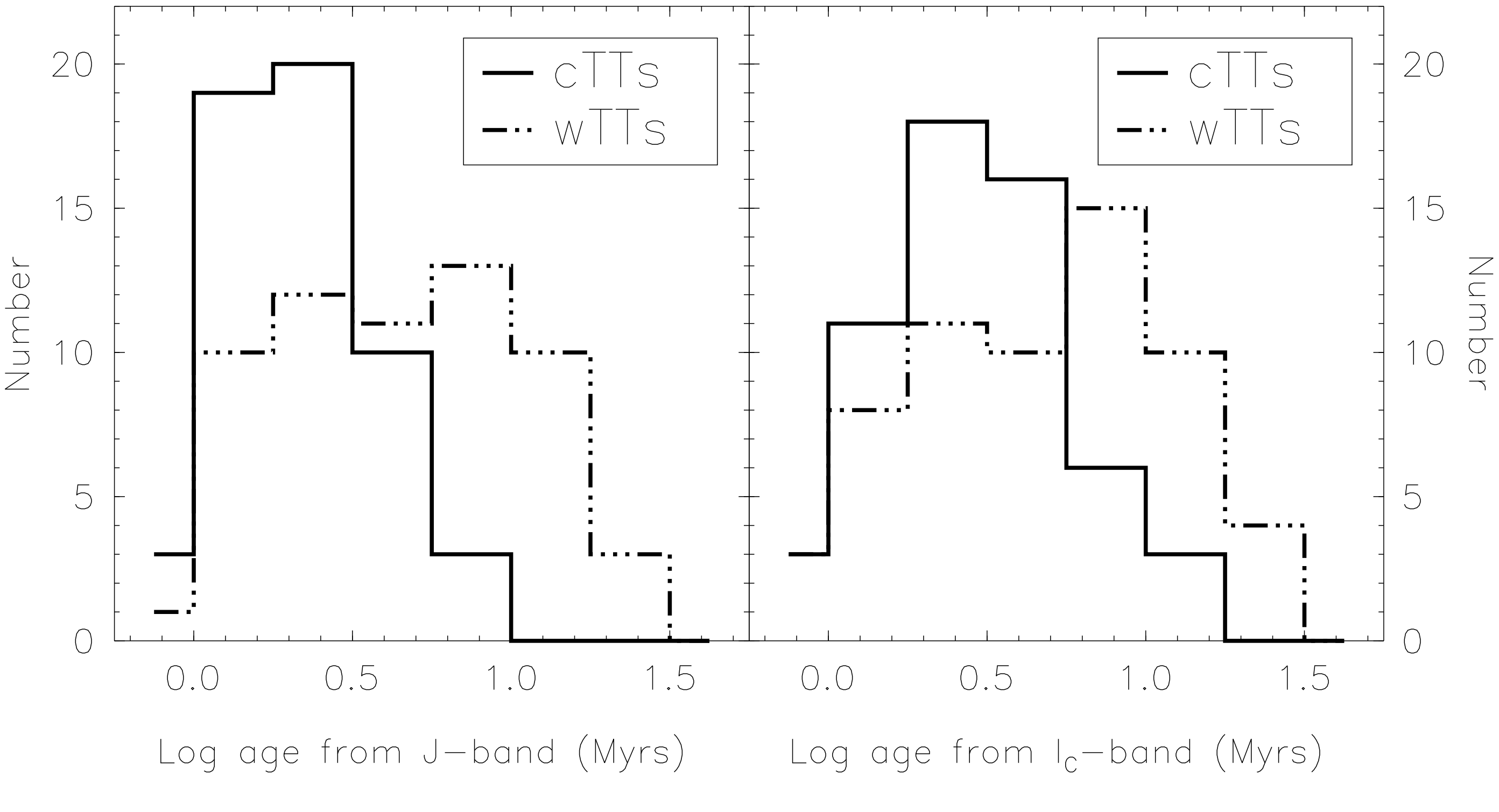